\documentclass[sigconf]{acmart}

\setcopyright{none}
\acmYear{2027}
\copyrightyear{2027}
\renewcommand\footnotetextcopyrightpermission[1]{}
\acmConference[KDD 2027]{}{}{}
\acmBooktitle{}
\acmDOI{}
\acmISBN{}

\usepackage{amsmath}
\usepackage{booktabs}
\usepackage{array}
\usepackage{multirow}
\usepackage{tabularx}
\usepackage{graphicx}
\usepackage{adjustbox}
\usepackage{algorithm}
\usepackage{algorithmic}
\usepackage{placeins}

\makeatletter
\def\@authorfont{\large\normalfont}
\def\@affiliationfont{\footnotesize\normalfont}
\makeatother

\newcommand{\method}{MM-ARC}
\newcommand{\rabo}{RABO}

\newcommand{\R}{\mathbb{R}}
\newcommand{\simplex}{\Delta}

\newcommand{\spaP}{.039}
\newcommand{\realityP}{.021}
\newcommand{\marketCostFullReturn}{13.9}
\newcommand{\marketCostFullSharpe}{1.31}
\newcommand{\marketCostLlmoeSharpe}{0.50}

\begin{document}

\title[MM-ARC: Adaptive Routing of Capital]{MM-ARC: Multimodal Adaptive Routing of Capital with Robustness-Audited Strategy Pools}

\author{Yang Chen}
\affiliation{%
  \institution{Zhejiang University}
  \city{Hangzhou}
  \country{China}}
\email{ychen2014@zju.edu.cn}

\author{Yuchen Cao}
\affiliation{%
  \institution{City University of Hong Kong}
  \city{Hong Kong}
  \country{China}}
\email{cynthcao2-c@my.cityu.edu.hk}

\author{Jacky Keung}
\affiliation{%
  \institution{City University of Hong Kong}
  \city{Hong Kong}
  \country{China}}
\email{jacky.keung@cityu.edu.hk}

\author{Leilei Gan}
\affiliation{%
  \institution{Zhejiang University}
  \city{Hangzhou}
  \country{China}}
\email{leileigan@zju.edu.cn}

\author{Kun Kuang}
\affiliation{%
  \institution{Zhejiang University}
  \city{Hangzhou}
  \country{China}}
\email{kunkuang@zju.edu.cn}

\author{Yueheng Jiang}
\affiliation{%
  \institution{Zhejiang University}
  \city{Hangzhou}
  \country{China}}
\email{yuehengj@zju.edu.cn}

\author{Zhaozhao Ma}
\affiliation{%
  \institution{Zhejiang University}
  \city{Hangzhou}
  \country{China}}
\email{zhaozhaoma@zju.edu.cn}

\author{Jianping Zhu}
\affiliation{%
  \institution{Dalian University of Technology}
  \city{Dalian}
  \country{China}}
\email{zhujp@mail.dlut.edu.cn}

\author{Fei Wu}
\affiliation{%
  \institution{Zhejiang University}
  \city{Hangzhou}
  \country{China}}
\email{wufei@zju.edu.cn}

\author{Jinpeng Li}
\authornote{Corresponding author: Jinpeng Li
  (\href{mailto:lijinpeng@zju.edu.cn}{lijinpeng@zju.edu.cn}).}
\affiliation{%
  \institution{Zhejiang University}
  \city{Hangzhou}
  \country{China}}
\email{lijinpeng@zju.edu.cn}

\renewcommand{\shortauthors}{Chen et al.}

\begin{abstract}
Financial trading systems must convert multimodal market history into executable positions while limiting overfitting from repeated strategy search. We introduce MM-ARC (MultiModal Adaptive Routing of Capital), which routes capital across trend, reversal, breakout, and exposure-control experts using aligned chart, numerical, and technical-text views. Within each market, regime-conditioned strategy pools are shared with bounded asset-specific adjustments. Robustness-Audited Bayesian Optimization (RABO) filters candidates proposed by Bayesian optimization on purged validation blocks using after-cost benchmark exceedance, lower-tail performance, stability, and turnover; a common portfolio layer then produces market-feasible orders. We evaluate 62 instruments across five asset classes using five training seeds and a frozen July 2025--June 2026 trading holdout. Under an all-in one-way cost of 10 basis points per unit of executed turnover, MM-ARC attains an equal-market Sharpe ratio of 1.33 and maximum drawdown of $-13.7\%$, versus 0.53 and $-18.3\%$ for the LLMoE-style routing baseline. The global learned-static control reaches 1.12 and $-15.3\%$, respectively. Paired block-bootstrap intervals favor the prespecified contrasts, while ablation point estimates are consistent with contributions from visual inputs, adaptive routing, exposure control, and robustness-audited admission. Family-level data-snooping tests also reject their prespecified nulls (SPA $p=\spaP$; Reality Check $p=\realityP$); we therefore interpret the evidence as benchmark-relative support within the evaluated candidate family and holdout, not as universal or future-regime superiority.
\end{abstract}



\maketitle
\fancyhead[LE]{}
\fancyhead[RO]{}

\section{Introduction}

Financial markets are non-stationary decision environments in which useful evidence is distributed across price sequences, chart structure, and textual market summaries. A model that predicts direction accurately can nevertheless fail as a trading system when its signals compete for capital, violate market constraints, or disappear after turnover and transaction costs. The central problem is therefore not perception alone, but the conversion of multimodal evidence into a synchronized, executable portfolio.

Financial language models and multimodal agents broaden the information interface of trading systems \cite{yang2023fingpt,xie2023pixiu,zhang2024finagent,timevlm2025}. Their forecasts and reasoning, however, neither allocate capital among incompatible trading behaviors nor reconcile their target exposures under a common portfolio budget. Ordinary mixture-of-experts routing selects representations or computation paths; a trading router must instead allocate capital, with direct consequences for exposure, turnover, and drawdown \cite{ding2024tradexpert,llmoe2025}.

Adaptation creates an opposing risk. A fixed strategy pool is too rigid under regime shifts, but selecting the best Bayesian optimization (BO) backtest or accepting unconstrained LLM-generated strategies enlarges the effective hypothesis space and amplifies data snooping \cite{white2000reality,hansen2005spa,bailey2014pbo,bailey2014deflated}. A credible adaptive pool must therefore use only historically visible information and judge candidates by the distribution---including lower-tail and drawdown behavior---rather than the peak of their validation outcomes.

We propose \method{}, an execution-aware pipeline that connects multimodal perception, adaptive capital routing, interpretable strategy experts, robustness-audited search, and market-feasible portfolio construction. We evaluate the complete pipeline on point-in-time records for 62 instruments from five asset classes using a one-year frozen trading holdout and a common after-cost portfolio-construction layer. The paper makes three contributions:
\begin{itemize}
  \item \textbf{Adaptive multimodal capital routing.} A shared encoder fuses aligned views, and a regime-conditioned router allocates capital over interpretable trend, reversal, breakout, and exposure-control experts before market-feasibility projection.
  \item \textbf{Market-shared expert strategy pools.} Instead of 744 asset-specific pools, \method{} uses 60 pools indexed by market, expert, and regime. Bounded, schema-valid asset-specific parameter adjustments (residuals) preserve strategy semantics and reduce repeated search.
  \item \textbf{Robustness-audited admission.} \rabo{} combines benchmark exceedance, validation lower-tail performance, median behavior, local parameter stability, and turnover. It thereby trades validation peaks for stronger lower-tail and drawdown behavior.
\end{itemize}

Across five separately settled market portfolios under the normalized all-in 10-bps protocol, the arithmetic means of SR and MDD are 1.33 and $-13.7\%$, versus .53 and $-18.3\%$ for the closest routing baseline. These are favorable holdout point differences of .80 in SR and 4.6 percentage points in MDD. SPA ($p=\spaP$) and Reality Check ($p=\realityP$) also reject their prespecified family-level nulls, providing benchmark-relative evidence after data-snooping correction on this holdout without implying dominance over every candidate or future regimes.

\section{Related Work}

\textbf{Financial LLM agents.} Financial LLMs and agents support text understanding, memory, forecasting, and multi-step trading workflows \cite{yang2023fingpt,wu2023bloomberggpt,xie2023pixiu,zhang2024finagent,yu2024finmem,xiao2024tradingagents}. \method{} addresses a narrower portfolio problem: translating multimodal states into capital weights, target exposures, and feasible orders rather than producing open-ended analyses or executable code.

\textbf{Multimodal forecasting.} Financial models increasingly combine structured series, images, and language, while Time-LLM and Time-VLM-style methods adapt foundation models to temporal prediction \cite{lee2019multimodal,jin2024time,timevlm2025}. Unlike systems that fuse independent news, filings, or macroeconomic text, \method{} uses visual and structured-text representations derived causally from the same observed OHLCV window and evaluates them by after-cost portfolio performance rather than forecasting accuracy alone.

\textbf{Routing and expert trading.} Modular trading systems and MoE models encourage specialization, but their routers commonly direct representations or select predictors \cite{ding2024tradexpert,llmoe2025}. In \method{}, the router instead produces simplex-valued capital allocations over interpretable strategy experts and is evaluated through after-cost portfolio contributions, turnover, exposure, and regime-conditioned outcomes.

\textbf{Backtest overfitting and robust search.} Reality Check, Superior Predictive Ability (SPA), Probability of Backtest Overfitting (PBO), Deflated Sharpe Ratio, and System Parameter Permutation diagnose selection risk; Bayesian optimization and LABO guide search \cite{white2000reality,hansen2005spa,bailey2014pbo,bailey2014deflated,walton2014know,pardo2008evaluation,snoek2012practical,chen2026labo}. \rabo{} embeds past-only cross-block distribution audits in admission and capital weighting.

\section{Problem Formulation}
\label{sec:problem}

Let $m\in\mathcal{M}$ index a market and $u\in\mathcal{U}_m$ an asset. At decision time $t$, the model encodes only observations available by that cutoff:
\begin{equation}
  o_{t,u}=(V_{t,u},X_{t,u},L_{t,u}),\qquad
  z_{t,u}=\operatorname{Enc}_{\eta}(o_{t,u}),
  \label{eq:encoding}
\end{equation}
where the chart, temporal, and structured-text views share the same timestamp. This cutoff-aligned state prevents later records from entering the decision.

The capital router converts state into continuous expert weights. It combines $z_{t,u}$, market $m$, the supervised fine-tuning (SFT) prior $p_{t,u}$, and pool evidence $c^{\rm rob}_{m,\hat r_{t,u}}$, where $\hat r_{t,u}=\arg\max_r p_{t,u,r}$ and $\eta_g\ge0$ controls the evidence tilt:
\begin{equation}
  w_{t,u}=\operatorname{softmax}\!\left(f_{\phi}(z_{t,u},m)+B_m p_{t,u}+\eta_g c^{\rm rob}_{m,\hat r_{t,u}}\right)
  \in\simplex^{K},
  \label{eq:router}
\end{equation}
with $K=4$. The simplex makes routing a capital reallocation among experts rather than a hard expert switch.

Every expert emits a target exposure $h_{t+1\mid t,u,k}\in[-1,1]$ in the same unit: a fraction of market-portfolio NAV before routing. To make these targets tradable, their router-weighted signals are synchronized on the native market calendar and projected onto the feasible set $\mathcal{C}_m$, which encodes position, leverage, short-sale, cash, and turnover limits:
\begin{subequations}
\begin{align}
  v_{t+1\mid t,m,k}&=\left(\frac{w_{t,u,k}h_{t+1\mid t,u,k}}{|\mathcal{U}_m|}\right)_{u\in\mathcal{U}_m},\notag\\
  h^{\rm raw}_{t+1\mid t,m}&=\sum_{k=1}^{K}v_{t+1\mid t,m,k},
  \label{eq:aggregation}\\
  h^{*}_{t+1\mid t,m}=\arg\min_{h\in\mathcal{C}_m}
  &\ \|h-h^{\rm raw}_{t+1\mid t,m}\|_2^2+\lambda_q\|h-h^{*}_{t,m}\|_1.
  \label{eq:projection}
\end{align}
\end{subequations}
Projection is performed separately for each market portfolio and reconciles both expert disagreement and cross-asset constraints. The order $q_{t+1,m}=h^*_{t+1\mid t,m}-h^*_{t,m}$ is submitted at the next tradable open, with cost and return booked to $t+1$; no expert receives credit for exposure removed by projection.

Settlement deducts applicable commissions, slippage, taxes, roll costs, funding, and soft feasibility penalties from realized next-period returns through $C_m(q)$. The learning target then discounts net returns and penalizes drawdown and volatility. All penalty weights are selected on validation records and frozen before the trading holdout:
\begin{subequations}
\begin{align}
r_{t+1,m}^{\rm net}
  &=\left(h^*_{t+1\mid t,m}\right)^{\!\top}r_{t+1,m}^{\rm asset}-C_m(q_{t+1,m}),
  \label{eq:net_return}\\
J_{\rm net}^{(m)}
  &=\mathbb{E}\!\left[\sum_{t=0}^{T}\gamma^t
  \left(r_{t+1,m}^{\rm net}-\lambda_{\rm dd}{\rm DD}_{t,m}
  -\lambda_{\rm vol}\sigma_{t,m}\right)\right].
  \label{eq:training_objective}
\end{align}
\end{subequations}
This closes the two-level decision process: asset-level observations determine expert targets, but only synchronized market holdings determine reward, risk, and the reported portfolio metrics.

\section{MM-ARC Framework}

\begin{figure*}[t]
  \centering
  \includegraphics[width=0.92\textwidth,height=0.613\textwidth,keepaspectratio=false]{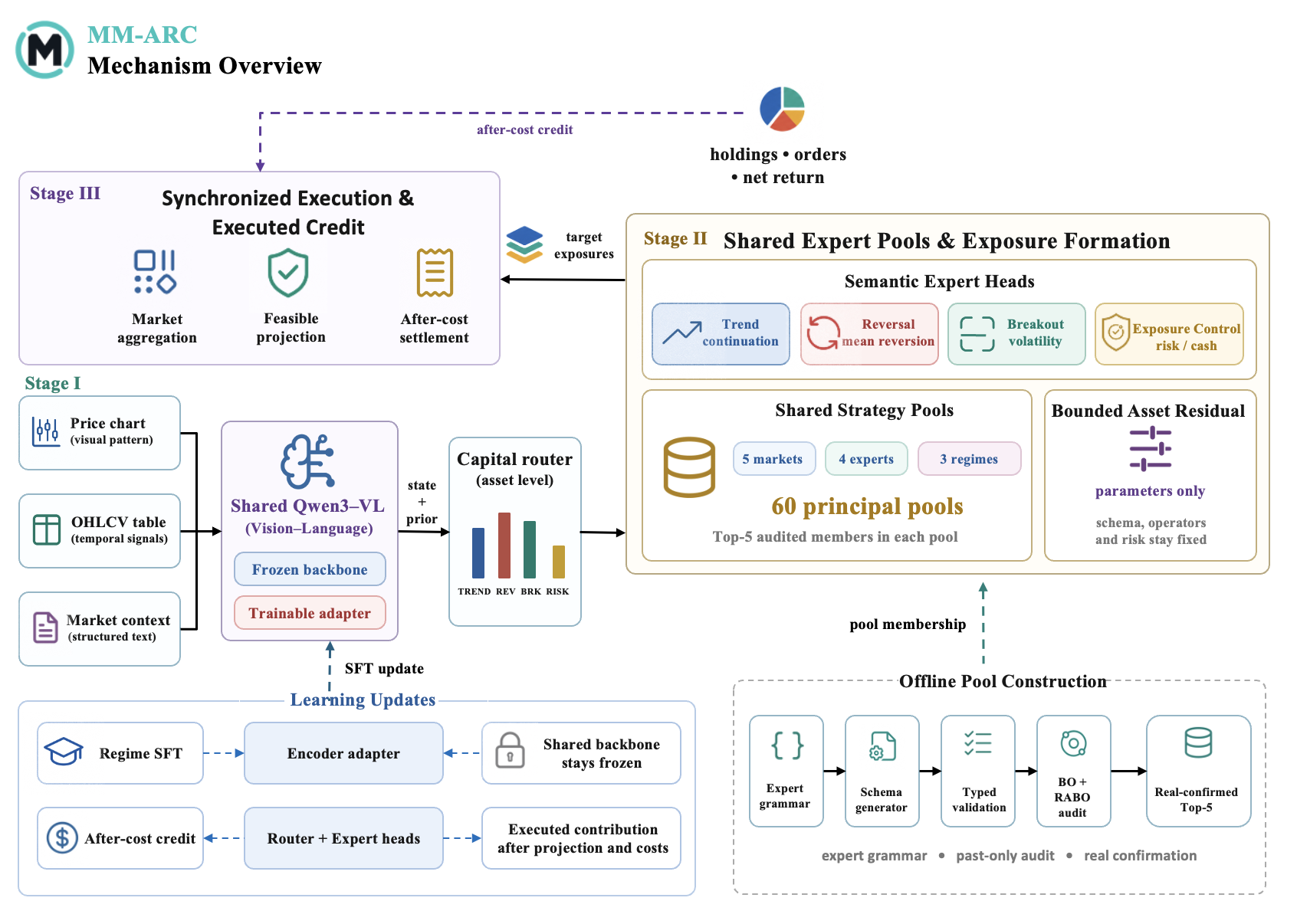}
  \Description{The MM-ARC mechanism maps chart, temporal, and structured-text inputs through a shared Qwen3-VL encoder and asset-level capital router to semantic expert heads, shared strategy pools with bounded asset residuals, synchronized feasible execution, and after-cost credit. Lower panels show offline pool construction and the learning updates.}
  \caption{Overview of \method{}. The solid path maps each asset's observed multimodal history to a fused state, continuous expert-capital weights, target exposures, and market-level portfolio construction. Dashed arrows denote robustness weighting and training feedback rather than online information flow.}
  \label{fig:framework}
\end{figure*}

\subsection{Multimodal Market Representation}

The representation stage asks what information is available at the decision cutoff and why multiple views are useful. Qwen3-VL-8B-Instruct encodes aligned chart, 100-observation numerical, and structured technical-summary views derived from the same OHLCV record \cite{bai2025qwen3vl}. These views expose visual patterns, numerical scale, and technical semantics without introducing an independent information source.

We use 4-bit QLoRA for parameter-efficient adaptation \cite{hu2021lora,dettmers2023qlora}. Supervised regime learning produces the prior $p_{t,u}$ from training records; its labels are not deployment inputs. The encoder, schema generator, and optional fidelity evaluator share only the base checkpoint, not adapters, prompts, output heads, or calibration records. Appendix~\ref{app:temporal} specifies view construction and label timing, and Appendix~\ref{app:qwen_roles} documents task isolation.

\subsection{Adaptive Capital Routing}

The router determines how much capital each interpretable expert receives. Equation~\eqref{eq:router} combines the fused market state with the SFT prior and the admitted pools' robustness evidence, producing continuous simplex weights rather than selecting one expert. The prior stabilizes state conditioning, while lower-tail evidence tilts capital toward pools that remain reliable on past validation blocks.

After SFT, the backbone is frozen and the adapters, router, and expert heads are optimized with post-projection, after-cost actor--critic reward. Entropy regularization discourages collapse, a prior penalty stabilizes the transition from SFT, and a weight-turnover penalty discourages unnecessary reallocations. Five independent runs use seeds 42--46, with checkpoints selected on validation records. Full training and executed-credit definitions appear in Appendix~\ref{app:training}.

\subsection{Expert Strategy Pools and Asset Adaptation}

The four experts encode trend, reversal, breakout, and exposure control. Candidates use typed, expert-specific schemas instead of free-form executable code; Appendix~\ref{app:shared_pools} gives their grammar and validation fields.

For market $m$, expert $k$, and regime $r$, the shared market--expert--regime strategy pool is
\begin{equation}
  \mathcal{P}_{m,k,r}=\{(s_j,\bar\theta_j,c^{\rm rob}_j)\}_{j=1}^{M_{\rm pool}},
  \qquad M_{\rm pool}=5.
  \label{eq:shared_pool}
\end{equation}
Five markets, four experts, and three regimes yield 60 shared pools rather than $62\times4\times3=744$. For asset $u$, the bounded asset-specific parameter adjustment (residual) is
\begin{equation}
  \theta_{u,j}=\Pi_{\Theta_{s_j}}\!\left(\bar\theta_j+D_{s_j}\delta_{u,j}\right),
  \quad \|\delta_{u,j}\|_{\infty}\le\epsilon_{\theta},
  \label{eq:residual}
\end{equation}
It can change only active numeric parameters, never operators, entry/exit logic, or risk constraints. Zero-centered regularization and declared bounds preserve shared strategy semantics; fitting uses only training data or historically visible validation blocks.

\subsection{Robustness-Audited Strategy Admission}

Admission asks whether a candidate is reliable across past validation blocks, not whether it attains the single highest validation return. Each candidate $x=(s,\theta)$ is therefore evaluated on synchronized tradable market portfolios:
\begin{equation}
\begin{aligned}
J_{\rm pool}(x;\mathcal{B}_b,m,k,r)
  &=J_{\rm net}\!\left(\operatorname{Port}_m(x,k,r;\mathcal{B}_b)\right),\\
D_{m,k,r}(x)
  &=\{J_{\rm pool}(x;\mathcal{B}_b,m,k,r)\}_{b=1}^{B_{m,r}}.
\end{aligned}
\label{eq:pool_objective}
\end{equation}
Here $\operatorname{Port}_m$ is the tradable market portfolio after applying the common projection and cost ledger. Appendix~\ref{app:shared_pools} gives the leave-one-asset influence diagnostic and sparse-cell shrinkage rule.

Bayesian optimization proposes schema--parameter pairs within the declared grammar. \rabo{} evaluates their past-only distributions and combines five within-update ranks:
\begin{equation}
 S_{\rm RABO}(x_i)=.30R_P+.30R_{Q^{\rm cand,val}_5}+.15R_{\rm med}+.15R_{\rm stab}-.10R_{\rm turn},
 \label{eq:rabo}
\end{equation}
The terms rank matched-block benchmark exceedance, the validation 5\% quantile, median behavior, local parameter stability, and realized turnover. The five highest audited candidates enter the pool. Appendix~\ref{app:rabo_details} provides the reproducible benchmark, stability neighborhood, controls, and DF-RABO efficiency study.

Selection and routing use related but distinct quantities. Equation~\eqref{eq:rabo} ranks candidates within one update; after admission, the pool-level robustness score for expert $k$ is
\begin{equation}
c^{\rm rob}_{m,r,k}=\sigma\!\left(\frac{Q_5(D^{\rm pool}_{m,k,r})-q_{\min}}{\tau_g}\right)
P_{D^{\rm pool}_{m,k,r}}(J_{\rm net}>J_{\rm bench})S^{\rm pool}_{m,k,r},
\label{eq:robust_gate}
\end{equation}
Here $J_{\rm bench}$ is the matched-block after-cost score of the predeclared fixed-seed strategy reference for the same market, expert, regime, and validation block. The reference uses the same synchronized portfolio construction and cost ledger, is fixed before BO and \rabo{}, and is never reselected candidate by candidate. $S^{\rm pool}$ is mean local stability, while $q_{\min}$ and $\tau_g$ are validation-frozen. Thus an admitted pool with deteriorating lower-tail evidence is down-weighted rather than treated as equally reliable.

\begin{figure*}[t]
  \centering
  \includegraphics[width=0.99\textwidth,keepaspectratio]{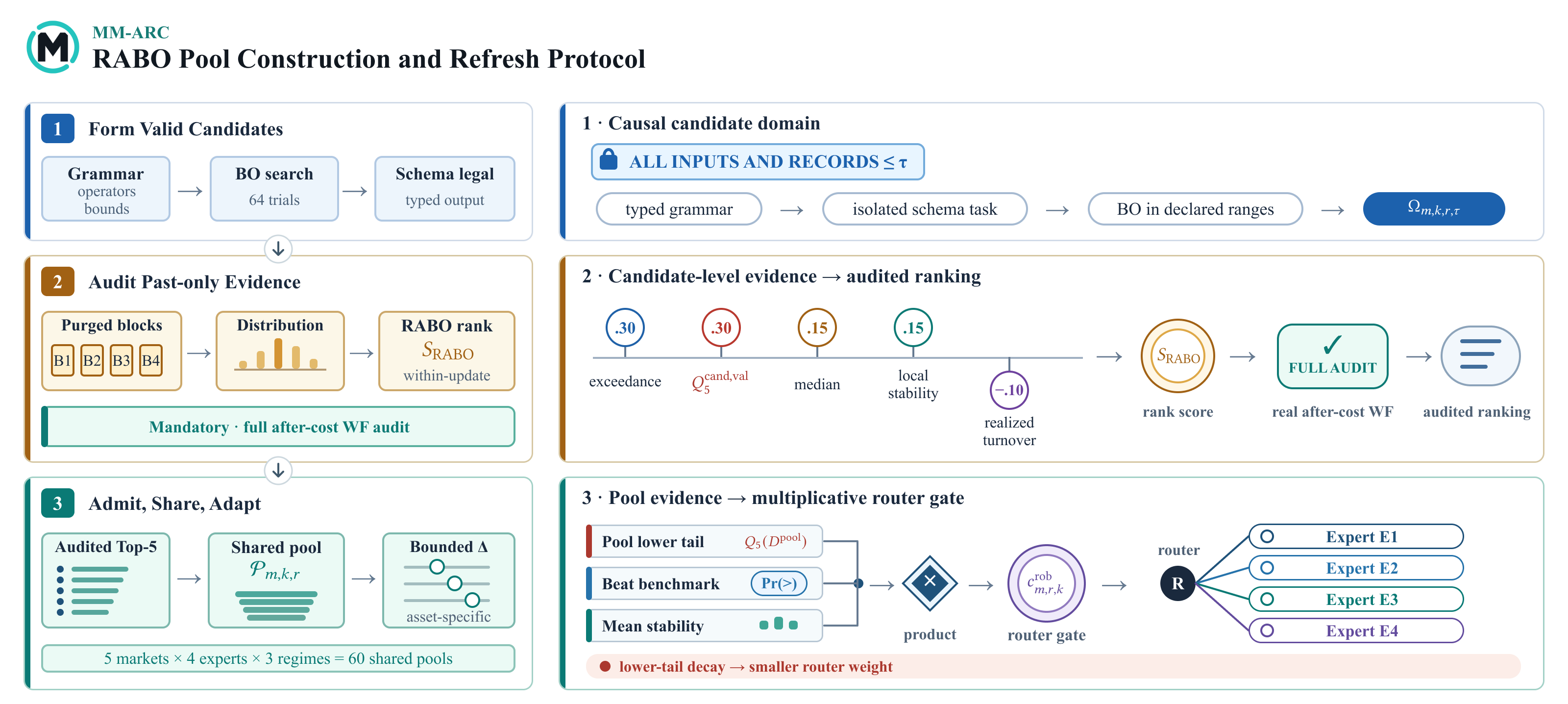}
  \Description{The three-stage RABO mechanism uses only records available by cutoff tau. It forms legal candidates through typed grammars and Bayesian optimization, ranks five robustness and turnover statistics on purged blocks before full after-cost walk-forward admission to a top-five shared pool with bounded asset residuals, and converts pool lower-tail, benchmark-exceedance, and stability evidence into a multiplicative gate on four expert router weights.}
  \caption{RABO pool construction and robustness-gated routing. At causal cutoff $\tau$, typed grammars and BO generate legal candidates; purged past-only blocks provide the five ranks in $S_{\rm RABO}$, and a full after-cost walk-forward audit admits the top five to shared market--expert--regime pools with bounded asset residuals. Pool lower-tail quality, benchmark exceedance, and mean stability are multiplied into $c^{\rm rob}_{m,r,k}$, which scales the corresponding expert's router weight.}
  \label{fig:rabo}
\end{figure*}

Under Online-WF, pool updates occur every 20 trading days or after a predeclared persistent-regime or validation-decay trigger. Every update at $\tau$ uses only observations through $\tau$. During the frozen trading holdout used for the main tables, schema generation, search, pool membership, adapters, router, and expert heads are all fixed before the holdout boundary.

\subsection{Market-Feasible Portfolio Construction}

Expert targets become orders only after synchronized, market-specific portfolio construction. Equations~\eqref{eq:aggregation}--\eqref{eq:projection} aggregate the asset-level expert targets and project them onto position, leverage, short-sale, cash, and turnover constraints. All trainable methods and compatible baselines use this same layer; A-shares additionally impose long-only feasibility.

At decision time $t$, the system uses records available through the market close and submits the projected order at the next tradable open. Equation~\eqref{eq:net_return} settles realized holdings and applicable costs, so reward and reported performance depend only on executed portfolios. Training attribution uses post-projection exposure and realized turnover cost; exact reconciliation, leave-one-expert-out diagnostics, the QP, and priority clipping are specified in Appendices~\ref{app:training} and \ref{app:execution}.

\section{Experimental Setup}
\label{sec:experiments}

We organize the evaluation around four questions: \textbf{RQ1}, how \method{}'s equal-market after-cost risk--return performance compares with the baselines, together with market-level heterogeneity relative to the passive benchmark; \textbf{RQ2}, whether adaptive capital routing, aligned modalities, and expert specialization contribute beyond matched controls; \textbf{RQ3}, whether \rabo{} improves lower-tail behavior relative to peak-based BO selection; and \textbf{RQ4}, whether market-shared pools reduce measured search workload without losing portfolio quality.

\subsection{Observed Market Data and Temporal Protocol}

The benchmark contains 62 real instruments: 10 A-shares, 15 U.S. equities, 15 ETFs, 20 futures, and Bitcoin and Ethereum. Core-market histories begin in January 2017; cryptocurrency histories begin in January 2021; all records end on June 30, 2026. The temporal modality contains OHLCV and technical indicators, the visual modality contains charts rendered from the same causal window, and the text modality contains deterministic technical summaries, regime descriptions, support/resistance levels, and volume analysis. Asset-specific calendars are retained.

The split follows one global calendar. Training ends in December 2023, validation spans January 2024 through June 2025, and the 90-observation labeling horizon is purged at split boundaries. Table~\ref{tab:temporal_protocol} distinguishes the one-year frozen trading holdout from the shorter regime-label evaluation. Throughout the holdout, all learned and selected states remain fixed; only forward inference, order generation, and ex post scoring are permitted. Regime-classification metrics end on January 31, 2026 (January 30 for exchange-traded sessions) because they require a 90-session forward label. Trading evaluation requires only next-period settlement and therefore continues through June 30, 2026. Every input, supervisory label, and candidate score carries a cutoff timestamp, and any source time later than its decision time is rejected.

\begin{table*}[t]
\centering
\caption{Temporal protocol and observed multi-market benchmark. Session counts cover the complete July 2025--June 2026 frozen trading holdout for each training seed.}
\label{tab:dataset_contract}
\label{tab:temporal_protocol}
\scriptsize
\setlength{\tabcolsep}{4.0pt}
\renewcommand{\arraystretch}{1.08}
\begin{tabularx}{\textwidth}{@{}l l >{\raggedright\arraybackslash}X@{}}
\toprule
\textbf{A. Phase} & \textbf{Dates} & \textbf{Permitted use} \\
\midrule
Training & Core: Jan. 2017--Dec. 2023; crypto: Jan. 2021--Dec. 2023 & Model and base-strategy fitting \\
Validation & Jan. 2024--Jun. 2025 & Checkpoint, RABO-weight, Top-$K$, refresh-threshold, and model-hyperparameter selection; 90-observation boundary purge \\
Frozen trading holdout & Jul. 1, 2025--Jun. 30, 2026 & Forward inference, orders, and ex post scoring only; all learned and selected states fixed \\
Regime-label evaluation & Jul. 1, 2025--Jan. 31, 2026 & Classification metrics requiring the 90-session forward label; Jan. 30 is the last exchange-traded session \\
Online-WF (supplementary) & Same initial state and trading period & Past-only causal updates; reported separately from the holdout result \\
\midrule
\end{tabularx}
\vspace{2pt}
\begin{tabularx}{\textwidth}{@{}l r r r c >{\raggedright\arraybackslash}X@{}}
\textbf{B. Market} & \textbf{Instruments} & \textbf{Sessions/seed} & \textbf{Annualization} & \textbf{Exposure support} & \textbf{Applicable cost components} \\
\midrule
A-shares & 10 & 242 & 252 & $[0,.95]$ & commission, slippage, sell-side stamp duty \\
U.S. equities & 15 & 251 & 252 & $[-.95,.95]$ & commission, slippage \\
ETFs & 15 & 251 & 252 & $[-.95,.95]$ & commission, slippage \\
Futures & 20 & 251 & 252 & $[-.95,.95]$ & commission, slippage, roll \\
Cryptocurrency & 2 & 365 & 365 & $[-.95,.95]$ & commission, slippage, funding \\
\bottomrule
\end{tabularx}
\end{table*}

The normalized 10-bps and market-specific 18/8/6/5/25-bps cost protocols are fixed before model selection and are never selected using validation performance.

Corporate actions are handled before feature construction; futures use lagged dominant-contract stitching and a separate roll ledger; cryptocurrency retains its 24/7 calendar. Appendix~\ref{app:temporal} reports data provenance, membership, missingness, and cutoff checks.

\subsection{Baselines, Training, and Fair Comparison}

The comparison covers Buy-and-Hold, a TPE-searched technical baseline, PPO, Time-VLM-style forecasting followed by the common allocation layer, FinAgent (adapted), and LLMoE-style routing \cite{haarnoja2018soft,schulman2017proximal,zhang2024finagent,timevlm2025,llmoe2025}. FinAgent and LLMoE-style are protocol-adapted local implementations because their native universes, information interfaces, and portfolio layers are not directly interchangeable with ours. This labeling avoids treating a local reproduction as the original paper's reported result.

The controls are chosen to isolate successive capabilities. Buy-and-Hold fixes the observed market path and supplies a passive reference. Stochastic technical search uses seeds 42--46 to initialize independent TPE-based BO runs with identical predeclared rules, parameter domains, validation objective, and fixed budget, then freezes each validation winner for the holdout; its variation thus reflects optimizer randomness, not different market histories. PPO tests continuous exposure control without the multimodal shared-pool architecture. Time-VLM-style maps aligned multimodal forecasts to target positions but lacks regime-conditioned shared pools. FinAgent (adapted) retains an agent-style decision interface under the common data and portfolio setting. LLMoE-style is the closest routing control: it uses four routed experts but removes the complete shared-pool and robustness-audit mechanism. The ablations additionally replace adaptive capital weights with uniform allocation, remove or corrupt one modality, flatten the expert structure, remove exposure control, and exchange \rabo{} for peak-, quantile-, or turnover-based admission.

To isolate state dependence from the benefit of merely learning a non-uniform expert mixture, we prespecify a learned-static-weight control for each training seed. It retains the same encoder, four experts, strategy pools, portfolio projection, cost ledger, validation objective, turnover constraint, and search budget as the full router, but optimizes one fixed simplex vector using validation records only:
\begin{equation}
  \begin{aligned}
    w_{\rm static}^{(s)}
    &=\arg\max_{w\in\simplex^{4}}J_{\rm val}^{(s)}(w),\\
    w_{t,u}^{(s)}
    &=w_{\rm static}^{(s)}
    \quad\text{for every holdout }t,u.
  \end{aligned}
  \label{eq:learned_static_control}
\end{equation}
During the holdout this control does not read $z_{t,u}$, the regime prior, or robustness-gate evidence. Its weights are never inferred from holdout averages. The adaptive-routing claim is evaluated against this control rather than against uniform weights or a single validation-best expert alone.

Every trainable method uses the same training/validation cutoffs, tradable universe, rebalance clock, and five seeds. Every method that emits a target position is passed through the same market-feasibility projection and cost accounting. Buy-and-Hold uses the common observed benchmark path and is therefore identical across training seeds. Checkpoint selection, BO/RABO weights, active-pool size, refresh thresholds, and all ablation settings are determined on training and validation records and frozen before the trading holdout.

All model and strategy selection ends before the holdout: checkpoints never use holdout targets, candidates are evaluated only on purged validation blocks, and reporting uses fixed prediction and order records. For each learned method, a seed identifies a complete independent training run rather than a resampling of the same return path. Five-seed dispersion therefore measures optimization stability, while the block bootstrap separately captures serial dependence in the held-out portfolio record.

\subsection{Execution, Metrics, and Statistical Inference}

All main results use one normalized, all-in charge on executed one-way turnover:
\begin{equation}
 C^{\rm norm}_{t,m}=c_{\rm norm}\lVert q_{t,m}\rVert_1,
 \qquad c_{\rm norm}=10^{-3}=10\ \text{bps},
 \label{eq:normalized_cost}
\end{equation}
where $\lVert q_{t,m}\rVert_1$ is the absolute executed change as a fraction of market-portfolio NAV. Hence 100\% one-way turnover costs 0.10\% of NAV; buying a fully invested position and later liquidating it costs approximately 0.20\% in total. Commission, slippage, sell-side stamp duty, roll cost, and funding fields allocate this single charge for attribution and are not added on top. For sensitivity analysis, predictions, gross returns, and orders are held fixed while $c_{\rm norm}$ changes over $\{0,5,10,20,50\}$ bps. The separate market-specific setting replaces Equation~\eqref{eq:normalized_cost}, rather than supplementing it, with effective rates of 18/8/6/5/25 bps for A-shares, U.S. equities, ETFs, futures, and cryptocurrency.

We compute total return (TR), Sharpe ratio (SR), maximum drawdown (MDD), annualized one-way turnover, and $Q^{\rm port}_{5,20d}$, the 5th percentile of rolling 20-session after-cost portfolio returns. The daily risk-free rate is zero; exchange-traded markets use 252 sessions per year and cryptocurrency uses 365. Annualized one-way turnover is $A_mT_m^{-1}\sum_t\lVert q_{t,m}\rVert_1$. Equations~\eqref{eq:sharpe_metric}--\eqref{eq:equal_market_metric} give the remaining ledger-level definitions and the equal-market reporting operator. Metrics are first computed for each seed--market portfolio and then equally averaged over the five markets. Mean and seed SD summarize five matched runs. The primary method-difference CI uses 1,200 draws from a 10-session market-stratified circular-block bootstrap applied to matched seed--market--time-block differences; method-level marginal CIs are retained in Appendix~\ref{app:numeric_diagnostics}. White's Reality Check and Hansen's SPA test address family-level data snooping. Holm-adjusted two-sided $t$-tests pair the five equal-market seed aggregates and are reported only as auxiliary optimization-stability evidence in the appendix.

The evidence criteria are fixed before reading the holdout results. RQ1 requires an equal-market comparison against all baselines together with explicit market-level heterogeneity and failure-case reporting for \method{} and the passive benchmark, rather than one favorable point estimate. RQ2 requires matched routing, modality, and expert controls with identical dates and settlement. RQ3 requires comparing BO and \rabo{} on the same portfolio-level lower-tail variable, together with return and drawdown. RQ4 requires completed-job counts and hardware logs, not pool-count arithmetic alone. A favorable paired test is not used to override an unfavorable SPA or Reality Check result.

Appendices~\ref{app:temporal} and \ref{app:result_provenance} document the evaluation protocol, ledger schemas, and reconciliation definitions.

\section{Results and Analysis}
\label{sec:results}

\subsection{Main Cross-Market Performance}

Table~\ref{tab:main_results} reports the one-year frozen trading holdout under the normalized all-in 10-bps protocol. Equal-market averages five separately settled portfolio metrics; MDD is not a pooled-curve drawdown. \method{} records TR 14.1\%, SR 1.33, and MDD $-13.7\%$, compared with 10.4\%, .53, and $-18.3\%$ for the closest routing baseline, LLMoE-style. The corresponding effect sizes are .80 in SR and 4.6 percentage points in MDD, while annualized one-way turnover falls from $12.5\times$ to $7.2\times$.

The market-level results remain heterogeneous. U.S. Buy-and-Hold has higher raw TR than \method{} (32.3\% versus 24.2\%), although \method{} records SR 1.55 and MDD $-16.5\%$ there. Cryptocurrency is \method{}'s weakest market portfolio (SR .94; MDD $-20.6\%$). Retaining this failure case in the equal-market aggregate makes the evidence about risk-adjusted performance and drawdown control rather than universal raw-return dominance.

\begin{table*}[t]
\centering
\caption{Equal-market results over the July 2025--June 2026 frozen trading holdout under the normalized 10-bps (0.10\%) one-way all-in cost. Values are mean $\pm$ sample SD over the five matched runs in Table~\ref{tab:per_seed_results}: training seeds for learned methods and TPE/BO seeds for stochastic technical search.}
\label{tab:main_results}
\scriptsize
\setlength{\tabcolsep}{8.0pt}
\begin{tabular}{lccc}
\toprule
Method & TR\% (Mean $\pm$ Seed SD) & SR (Mean $\pm$ Seed SD) & MDD\% (Mean $\pm$ Seed SD) \\
\midrule
B\&H & 14.0 $\pm$ 0.0 & 0.58 $\pm$ 0.00 & -24.7 $\pm$ 0.0 \\
Stoch. technical search & 8.9 $\pm$ 0.9 & 0.45 $\pm$ 0.03 & -20.3 $\pm$ 0.6 \\
PPO & 8.4 $\pm$ 0.6 & 0.45 $\pm$ 0.06 & -18.3 $\pm$ 0.3 \\
Time-VLM-style & 9.7 $\pm$ 0.9 & 0.49 $\pm$ 0.06 & -18.5 $\pm$ 0.5 \\
FinAgent (adapted) & 8.7 $\pm$ 0.8 & 0.44 $\pm$ 0.03 & -20.6 $\pm$ 0.3 \\
LLMoE-style & 10.4 $\pm$ 0.9 & 0.53 $\pm$ 0.05 & -18.3 $\pm$ 0.4 \\
\textbf{MM-ARC} & \textbf{14.1 $\pm$ 0.9} & \textbf{1.33 $\pm$ 0.06} & \textbf{-13.7 $\pm$ 0.2} \\
\bottomrule
\end{tabular}
\end{table*}

\begin{table*}[t]
\centering
\caption{Difference-focused evidence on the frozen holdout. Effects are the first-named method minus the control; positive $\Delta$MDD and $\Delta Q^{\rm port}_{5,20d}$ are favorable. The 95\% CIs are obtained by resampling matched seed--market--time-block differences. SPA and Reality Check are family-level diagnostics shown once.}
\label{tab:statistical_summary}
\scriptsize
\setlength{\tabcolsep}{3.2pt}
\renewcommand{\arraystretch}{1.04}
\resizebox{\textwidth}{!}{%
\begin{tabular}{lrrrrrrrr}
\toprule
Comparison
& $\Delta$SR & 95\% CI
& $\Delta$MDD (pp) & 95\% CI
& $\Delta Q^{\rm port}_{5,20d}$ (pp) & 95\% CI
& SPA $p$
& \shortstack{Reality\\Check $p$} \\
\midrule
MM-ARC $-$ LLMoE-style
& $+0.80$ & $[0.26,1.46]$
& $+4.6$ & $[2.03,6.94]$
& $+2.15$ & $[.78,3.56]$
& $\spaP$
& $\realityP$ \\
MM-ARC $-$ global learned static
& $+0.21$ & $[0.10,0.39]$
& $+1.6$ & $[.25,2.70]$
& $+1.38$ & $[.30,2.57]$
& ---
& --- \\
MM-ARC $-$ market-specific learned static
& $+0.12$ & $[.06,.21]$
& $+0.8$ & $[.05,1.32]$
& $+0.63$ & $[.15,1.14]$
& ---
& --- \\
Full RABO $-$ BO-best
& $+0.31$ & $[.13,.55]$
& $+3.6$ & $[1.36,5.68]$
& $+3.63$ & $[1.64,5.65]$
& ---
& --- \\
\bottomrule
\end{tabular}
}
\end{table*}

Table~\ref{tab:statistical_summary} reports matched method differences and their paired block-bootstrap intervals. Against LLMoE-style, $\Delta{\rm SR}=+.80$ with CI $[.26,1.46]$; the risk-effect point estimates are $\Delta{\rm MDD}=+4.6$ and $\Delta Q^{\rm port}_{5,20d}=+2.15$ percentage points. Against global learned-static weights, the corresponding differences are $+.21$, $+1.6$, and $+1.38$, with the SR CI $[.10,.39]$. Against market-specific learned-static weights, $\Delta{\rm SR}=+.12$ with CI $[.06,.21]$. Under the prespecified benchmark and candidate-family construction, SPA ($p=\spaP$) and Reality Check ($p=\realityP$) reject their family-level nulls. The family-level evidence therefore supports benchmark-relative superiority within the evaluated family and holdout, but not dominance over every candidate or future-regime generalization. Block-length results appear in Appendix~\ref{app:numeric_diagnostics}.

\subsection{Routing, Modalities, and Expert Structure}

\begin{table*}[t]
\centering
\caption{Five-seed mean ablations on the frozen holdout. Metrics are equal-market summaries and $Q^{\rm port}_{5,20d}$ is the common rolling-20-session lower-tail statistic. Global learned-static weights follow Eq.~\eqref{eq:learned_static_control}.}
\label{tab:combined_ablation}
\scriptsize
\setlength{\tabcolsep}{5.0pt}
\begin{tabular}{llrrrr}
\toprule
Panel & Variant & TR\% & SR & MDD\% & $Q^{\rm port}_{5,20d}$\% \\
\midrule
\multirow{5}{*}{A: Routing} & \textbf{Regime-tilted router} & \textbf{14.1} & \textbf{1.33} & \textbf{-13.7} & \textbf{-8.22} \\
 & Global learned static weights & 13.5 & 1.12 & -15.3 & -9.60 \\
 & Market-specific learned static & 13.8 & 1.21 & -14.5 & -8.85 \\
 & Uniform weights & 11.3 & 0.88 & -16.6 & -11.60 \\
 & Validation-best expert & 12.0 & 0.98 & -17.6 & -12.00 \\
\midrule
\multirow{3}{*}{B: Modalities} & \textbf{Full aligned views} & \textbf{14.1} & \textbf{1.33} & \textbf{-13.7} & \textbf{-8.22} \\
 & Temporal + text & 12.6 & 1.14 & -15.3 & -10.91 \\
 & Corrupted visual & 10.9 & 0.85 & -17.2 & -11.35 \\
\midrule
\multirow{3}{*}{C: Experts} & \textbf{Four semantic experts} & \textbf{14.1} & \textbf{1.33} & \textbf{-13.7} & \textbf{-8.22} \\
 & Flat pool & 11.1 & 0.90 & -17.3 & -11.62 \\
 & No exposure control & 13.0 & 1.10 & -18.2 & -13.64 \\
\midrule
\multirow{8}{*}{D: Pools/search} & Fixed seed pool & 10.2 & 0.79 & -18.4 & -11.61 \\
 & BO-best & 15.0 & 1.02 & -17.3 & -11.85 \\
 & Turnover-matched BO & 14.3 & 0.99 & -16.2 & -9.87 \\
 & Q5-only audit & 13.2 & 1.14 & -14.2 & -8.20 \\
 & Equal-weight audit & 13.6 & 1.20 & -14.5 & -9.59 \\
 & Asset-specific RABO & 13.9 & 1.26 & -13.9 & -8.57 \\
 & Shared, no residual & 13.4 & 1.22 & -14.2 & -9.36 \\
 & \textbf{Shared + residual (full)} & \textbf{14.1} & \textbf{1.33} & \textbf{-13.7} & \textbf{-8.22} \\
\bottomrule
\end{tabular}
\end{table*}

Panel A of Table~\ref{tab:combined_ablation} answers RQ2. The router records TR 14.1\%, SR 1.33, MDD $-13.7\%$, and Q5 $-8.22\%$; global static gives 13.5\%, 1.12, $-15.3\%$, and $-9.60\%$, while market-specific static gives 13.8\%, 1.21, $-14.5\%$, and $-8.85\%$. Uniform and validation-best weights reach SR .88 and .98; removing or corrupting the visual view lowers SR to 1.14 and .85.

Panel C evaluates expert decomposition. A flat pool reaches SR .90, and removing exposure control deepens MDD from $-13.7\%$ to $-18.2\%$. Thus, the exposure expert is not merely an interpretability label: it materially changes the executed risk profile. The benefit of the full four-expert model over the flat-pool control also supports the use of semantic restrictions during strategy search.

\subsection{RABO and Shared-Pool Analysis}

Panel D answers RQ3. BO-best delivers the highest TR (15.0\%) but has $Q^{\rm port}_{5,20d}=-11.85\%$ and MDD $-17.3\%$. Full \rabo{} gives $-8.22\%$ and $-13.7\%$ at TR 14.1\%. The Q5-only rule is marginally better on the designated quantile ($-8.20\%$), but lower on SR (1.14 versus 1.33) and worse on MDD ($-14.2\%$ versus $-13.7\%$). The composite audit therefore trades 0.9 percentage points of TR relative to the validation-peak selector for a broader improvement in lower-tail and drawdown behavior.

The shared-plus-residual design reaches SR 1.33, compared with 1.26 for asset-specific \rabo{} and 1.22 for sharing without residuals. Relative to asset-specific \rabo{}, its point differences are $+0.2$ percentage points in TR, $+.07$ in SR, $+0.2$ percentage points in MDD, and $+0.35$ percentage points in $Q^{\rm port}_{5,20d}$. Table~\ref{tab:shared_pool_difference} reports the matched-block intervals for these differences. All four intervals cross zero, so the comparison does not establish a quality difference between the shared and asset-specific designs. The observed pattern is consistent with market-level sharing providing common evidence while bounded asset-specific adjustments preserve useful heterogeneity; workload reduction is evaluated separately below.

\begin{table}[t]
\centering
\caption{Shared + residual minus asset-specific \rabo{}. Positive differences favor the shared design. Confidence intervals use the same 1,200-draw matched seed--market--time-block circular-block bootstrap as the primary comparisons.}
\label{tab:shared_pool_difference}
\scriptsize
\setlength{\tabcolsep}{4.0pt}
\begin{tabularx}{\columnwidth}{@{}X r c@{}}
\toprule
Metric & Point difference & Paired-block 95\% CI \\
\midrule
TR (percentage points) & $+0.2$ & $[-1.4,+2.0]$ \\
SR & $+0.07$ & $[-0.20,+0.37]$ \\
MDD (percentage points) & $+0.2$ & $[-1.3,+1.5]$ \\
$Q^{\rm port}_{5,20d}$ (percentage points) & $+0.35$ & $[-0.9,+1.5]$ \\
\bottomrule
\end{tabularx}
\end{table}

\subsection{Mechanism, Workload, and Cost Diagnostics}

\begin{figure*}[t]
  \centering
  \begin{minipage}[t]{0.47\textwidth}
    \centering\textbf{(a) Same-variable lower tail}\par\smallskip
    \includegraphics[width=\linewidth]{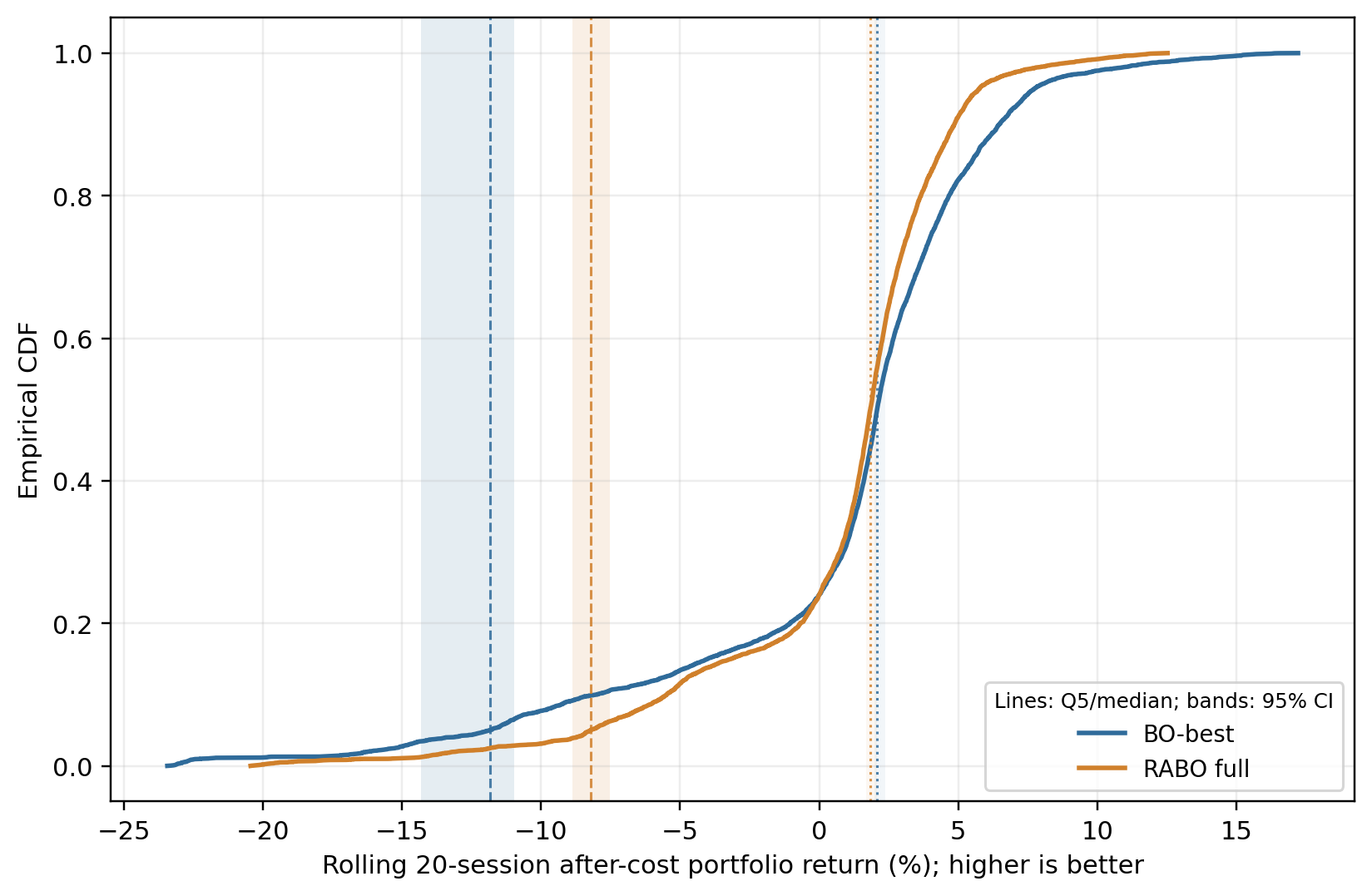}
  \end{minipage}\hfill
  \begin{minipage}[t]{0.47\textwidth}
    \centering\textbf{(b) Regime-Tilted Capital Weights}\par\smallskip
    \includegraphics[width=\linewidth]{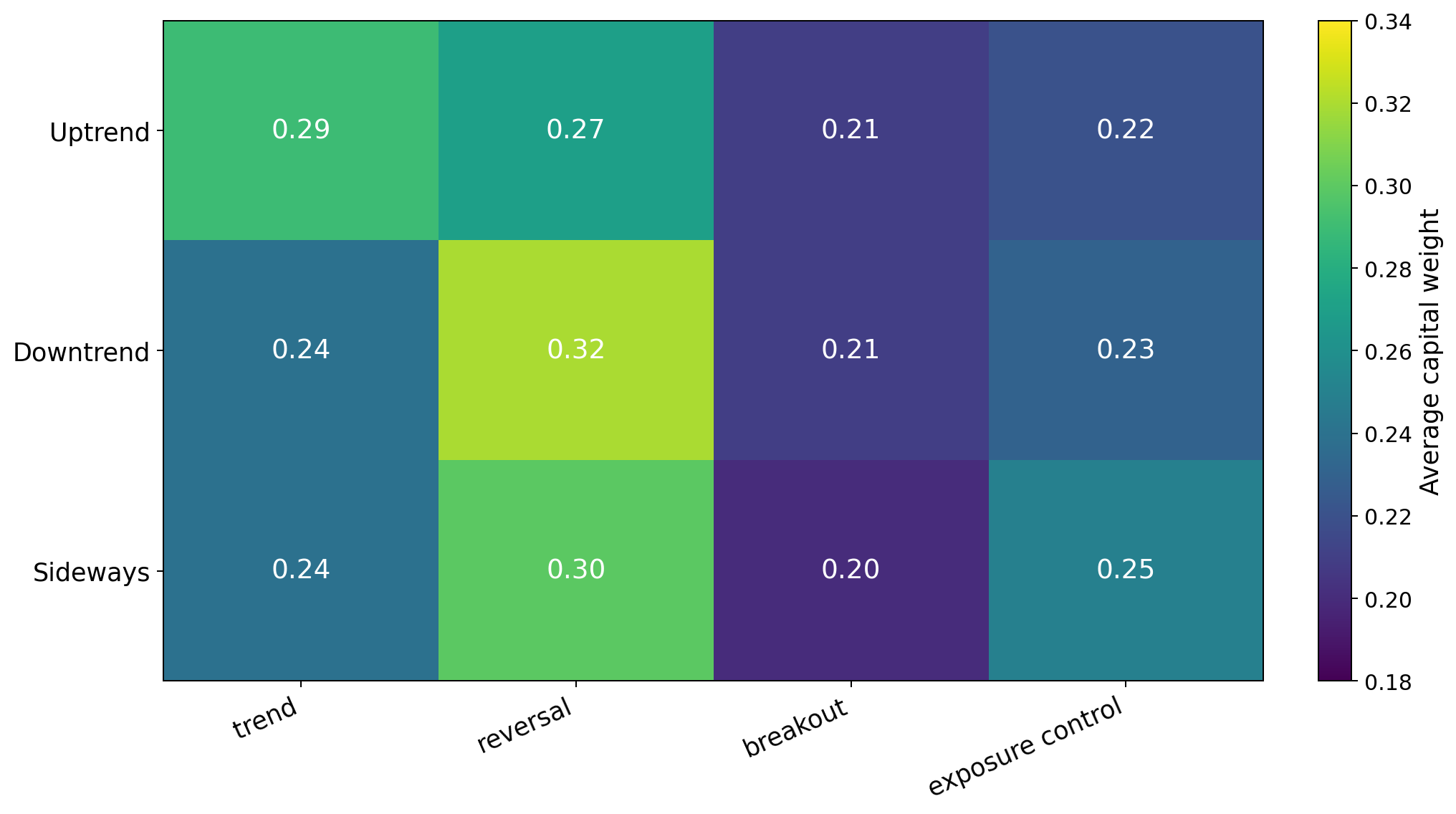}
  \end{minipage}
  \Description{The left panel compares empirical CDFs of the same rolling 20-session after-cost portfolio return for BO-best and full RABO. The right panel shows average capital weights for four experts under three predicted market regimes.}
  \caption{Mechanism diagnostics. Panel (a) marks Q5 with dashed lines and the median with dotted lines; shaded bands are 95\% seed--market cluster-bootstrap intervals. Panel (b) summarizes capital weights over the three regimes that index the shared pools.}
  \label{fig:mechanism}
\end{figure*}

Figure~\ref{fig:mechanism}(a) compares BO-best and \rabo{} on the same rolling after-cost portfolio variable. \rabo{} shifts Q5 rightward while leaving the median slightly lower, whereas BO-best retains higher full-period TR. In panel (b), trend peaks in uptrends (.29), reversal in downtrends (.32), and exposure control in sideways periods (.25). Temporal weight variance is .00192, pairwise symmetric KL divergence is .00677, and entropy is 1.363. These modest but systematic state-dependent tilts are consistent with adaptive capital routing within a diversified expert mixture, rather than hard expert selection or discrete regime switching.

\begin{table*}[t]
\centering
\caption{Measured search workload under a fixed hardware and software environment. Counts include completed jobs; time and peak memory come from the recorded resource logs.}
\label{tab:resource_results}
\scriptsize
\setlength{\tabcolsep}{4.0pt}
\begin{tabular}{lrrrrrrrrr}
\toprule
Design & Gen. & Valid & BO prop. & Backtests & Residual h & E2E h & CPU h & GPU h & Peak GB \\
\midrule
Asset-specific RABO & 4,464 & 3,578 & 47,616 & 43,192 & 0.00 & 31.6 & 126.4 & 18.2 & 28.4 \\
\textbf{Shared + residual} & \textbf{360} & \textbf{291} & \textbf{3,840} & \textbf{3,594} & 0.42 & \textbf{4.6} & \textbf{18.7} & \textbf{3.2} & \textbf{21.8} \\
\bottomrule
\end{tabular}
\vspace{2pt}

\parbox{0.98\textwidth}{\scriptsize
\textit{Measurement environment}. EPYC 7543P/256\,GB/A100 40\,GB;
Python 3.11, PyTorch 2.7.1/CUDA 12.8; four FIFO workers, batch 1, cache off,
failures/retries included; E2E includes all stages through table write-out.
One-second \texttt{psutil}/\texttt{nvidia-smi} sampling; three-run E2E ranges:
30.9--32.4\,h and 4.4--4.8\,h.}
\end{table*}

For RQ4, structural sharing reduces the pool count from 744 asset-specific pools to 60 market--expert--regime pools. The measured workload also falls: complete backtests decline from 43,192 to 3,594, and end-to-end time from 31.6 to 4.6 hours, including 0.42 hours of residual fitting. CPU time falls from 126.4 to 18.7 hours, GPU time from 18.2 to 3.2 hours, and peak memory from 28.4 to 21.8 GB under the same environment. The smaller proportional wall-clock gain relative to the pool-count reduction reflects fixed orchestration, data loading, and residual-fitting overheads.

Holding predictions, gross returns, and orders fixed, increasing the normalized all-in rate from 10 to 50 bps reduces \method{} TR from 14.1\% to 10.8\% and SR from 1.33 to 1.00; LLMoE-style falls from 10.4\% to 4.8\% and from .53 to .24. Under the replacement market-specific 18/8/6/5/25-bps schedule, \method{} records TR $\marketCostFullReturn\%$ and SR $\marketCostFullSharpe$, versus LLMoE-style SR $\marketCostLlmoeSharpe$. The qualitative comparison is therefore not tied to the normalized 10-bps point.

\subsection{Research-Question Synthesis}

The four analyses support a qualified conclusion. RQ1 favors \method{} in risk-adjusted point estimates and paired intervals, and both family-level tests reject their prespecified nulls, although the market-level results retain meaningful heterogeneity. RQ2's matched controls favor adaptive routing over both learned-static alternatives and support the modalities and architecture. RQ3 shows a 0.9-point TR trade-off for shallower MDD and a less negative Q5. RQ4 shows lower pool count and measured work. These findings connect architectural choices to executed outcomes without relying on one headline metric.

\section{Discussion and Limitations}

\textbf{Temporal scope.} The July 2025--June 2026 frozen trading holdout is one historical realization and cannot establish adaptation across longer market cycles. In the retained cryptocurrency failure, seed 46 loses 20.5\% between January 18 and February 5, 2026, showing that positive full-period TR and SR can coexist with a concentrated loss when a fast volatility transition outpaces the frozen pool. Longer forward paper trading and prospectively declared causal refresh checks are therefore required.

\textbf{Statistical scope.} SPA ($p=\spaP$) and Reality Check ($p=\realityP$) reject their family-level nulls under the prespecified benchmark, candidate set, and performance differential. These results strengthen the benchmark-relative evidence after data-snooping correction on the observed holdout, but they do not show that \method{} dominates every candidate or that the advantage generalizes to future regimes. Five seeds measure optimization variability, not uncertainty over alternative market histories; the block bootstrap captures dependence within the same holdout rather than variation across future regimes.

\textbf{Universe and information scope.} The 62 instruments across five asset classes do not represent the complete investable universe, and survivorship bias may remain where point-in-time membership is unavailable. The text view is derived from market records, so the study does not test independent news, filings, or macroeconomic reasoning. Larger and more heterogeneous universes may also require sector- or liquidity-level parents rather than one shared hierarchy per market.

\textbf{Execution and capacity scope.} The common ledger models commissions, slippage, stamp duty, roll cost, funding, position caps, and turnover, but not every queue, borrow, outage, capacity, or nonlinear market-impact effect. Capacity-aware simulation and forward execution are needed before interpreting the observed results as deployable scale.

\section{Conclusion}

On this frozen holdout, \method{} yields favorable risk-adjusted and drawdown estimates relative to the closest routing baseline and both learned-static controls, and SPA and Reality Check reject their prespecified family-level nulls. Retained ablations and workload measurements are consistent with contributions from adaptive routing, the visual view, exposure control, \rabo{}, and shared-pool workload reduction. Together, these results support a benchmark-relative advantage within the evaluated candidate family and period, without claiming dominance over every candidate or future-regime superiority.

\clearpage
\bibliographystyle{ACM-Reference-Format}
\bibliography{references}

\clearpage
\appendix
\begingroup
\setlength{\fboxsep}{6pt}
\noindent\fbox{%
  \parbox{\dimexpr\columnwidth-2\fboxsep-2\fboxrule\relax}{%
    \centering
    \textbf{Anonymous Executable Pipeline and Experiment Results}\\[2pt]
    \href{https://anonymous.4open.science/r/MM-ARC-32F7/}{%
      \textbf{Open repository}\,\raisebox{0.15ex}{\(\nearrow\)}}\\[1pt]
    {\scriptsize
      \href{https://anonymous.4open.science/r/MM-ARC-32F7/}{%
        \texttt{anonymous.4open.science/r/MM-ARC-32F7}}}
  }%
}
\endgroup

\section{Temporal Evaluation and Data Provenance}
\label{app:temporal}

\subsection{Calendar and Frozen Evaluation}

The benchmark is constructed from point-in-time market records. Core-market histories span January 2017 through June 30, 2026, and cryptocurrency histories span January 2021 through the same cutoff. Core-market training uses January 2017 through December 2023, cryptocurrency training begins in January 2021, and validation spans January 2024 through June 2025. The frozen trading holdout runs from July 1, 2025 through June 30, 2026. A-shares, U.S. equities, ETFs, and futures follow their native exchange calendars, while cryptocurrency uses a continuous daily calendar; returns are not forward-filled across closed markets.

Training, validation, BO, \rabo{}, schema formation, and checkpoint selection are completed before the trading holdout is opened. Throughout the full year, the encoder adapters, router, expert heads, shared pools, asset-specific parameter adjustments, and execution configuration remain fixed. Regime-classification metrics are reported through January 31, 2026 (January 30 for exchange-traded sessions) because their labels require 90 future sessions; after-cost trading metrics extend through June 30 because they require only next-period settlement. The supplementary Online-WF protocol begins from the same initial state, is reported separately, and restricts every update at $\tau$ to inputs, labels, validation blocks, and candidate evaluations available no later than $\tau$.

\begin{table*}[h]
\centering
\caption{Benchmark composition and operational handling. Every row is evaluated on observed market records under its native calendar.}
\label{tab:dataset_appendix}
\small
\begin{tabularx}{\textwidth}{lrrXX}
\toprule
Market & Instruments & Evaluation sessions/seed & Position and settlement & Data handling \\
\midrule
A-shares & 10 & 242 & long-only, T+1, sell-side stamp duty & adjusted prices, suspensions, exchange calendar \\
U.S. equities & 15 & 251 & bounded long/short, margin limits & corporate actions, borrow and exchange-calendar checks \\
ETFs & 15 & 251 & venue-dependent shorting and cash limits & distributions and underlying-asset classification \\
Futures & 20 & 251 & contract leverage, margin, next-open rolls & lagged dominant contract and explicit roll ledger \\
Cryptocurrency & 2 & 365 & continuous trading and bounded leverage & exchange-source, funding, and outage checks \\
\midrule
Total & 62 & --- & five separately settled market portfolios & equal-market reporting after portfolio-level metrics \\
\bottomrule
\end{tabularx}
\end{table*}

\subsection{Artifact Integrity and Leakage Controls}

The executable pipeline applies provenance and leakage checks to its inputs and processed samples. For each available input artifact, the validator records its source, retrieval timestamp, adjustment convention, first and last observation, row count, missingness, duplicate count, and SHA-256 digest. Processed multimodal samples additionally carry the instrument, decision timestamp, maximum source timestamp, feature version, chart-rendering version, and split identifier. A sample fails validation if any source timestamp is later than its decision time or if its target timestamp is visible to training, schema formation, BO, \rabo{}, or checkpoint selection.

Portfolio ledgers produced by the pipeline carry method, training seed, market, date, protocol phase, previous holding, requested target, executed holding, order, rejected notional, gross return, component costs, and net return. Instrument records retain stable market identifiers and point-in-time universe membership. These schemas support record-level reconciliation of predictions, orders, holdings, costs, and reported metrics during pipeline execution.

\subsection{Multimodal View and Label Construction}

At each decision time, the sample builder extracts the 100 observations ending at the recorded cutoff. The visual branch renders a standard candlestick chart and a CYC-MRKAB chart containing moving averages, Bollinger Bands, MACD, RSI, and KDJ. The temporal branch contains OHLCV and technical indicators normalized with training-only statistics. The structured-text branch deterministically summarizes returns, volatility, moving-average ordering, indicator positions, support/resistance, and volume from the same window. The three branches are generated separately and then joined by instrument and decision time; none adds independent news, filings, or macroeconomic records.

SFT targets use the designated 90-session forward regime horizon only within the labeling phase. Samples whose target horizon crosses a split boundary are purged, and target timestamps are inaccessible to feature construction, strategy search, and checkpoint selection. These labels supervise and evaluate the regime prior $p_{t,u}$ but are never supplied as contemporaneous inference inputs.

\section{Qwen3-VL Roles and Training Isolation}
\label{app:qwen_roles}

The implementation uses three task-separated Qwen3-VL roles. They share the Qwen3-VL-8B-Instruct base checkpoint but not adapters, output heads, prompts, or task-specific training state. The execution evaluator is deterministic and does not use a language-model judgment.

\begin{table*}[h]
\centering
\caption{Task separation in the implemented model. Only full walk-forward evaluation can authorize pool admission.}
\label{tab:qwen_roles}
\small
\begin{tabularx}{\textwidth}{lXXX}
\toprule
Role & Input/output & Isolation & Authority \\
\midrule
Multimodal encoder & charts, series, summary $\rightarrow z_{t,u},p_{t,u}$ & encoder adapter and regime head & representation and routing only \\
Schema generator & grammar and past context $\rightarrow$ typed schema & generator adapter, prompt, and decoding seed & candidate proposal only \\
Fidelity evaluator & schema and past context $\rightarrow J_L$ & evaluator adapter and calibrated regression head & acquisition priority only \\
Execution evaluator & executable schema $\rightarrow J_R$ & fixed walk-forward engine and validation protocol & mandatory evidence for admission \\
\bottomrule
\end{tabularx}
\end{table*}

The encoder is first adapted with regime supervision and then used by the after-cost actor--critic stage. The generator is trained to emit a typed schema that passes expert-specific grammar, indicator, parameter-range, and risk-control checks. The fidelity evaluator is calibrated only on past paired evaluations. No evaluator score can substitute for $J_R$ when a candidate enters the Top-5 active pool.

\section{Shared Pools and Search-Space Accounting}
\label{app:shared_pools}

\subsection{Pool Hierarchy}

The shared pool $\mathcal{P}_{m,k,r}$ reuses strategy structure and base parameters for one market, expert type, and regime. Equation~\eqref{eq:residual} limits each asset to bounded, schema-valid continuous or integer offsets (residuals). Categorical operators, entry/exit logic, and risk constraints remain shared. Residual clipping uses the tighter of the schema range and a predeclared relative radius; zero-centered initialization and an $\ell_2$ penalty discourage unnecessary personalization.

\begin{table}[t]
\centering
\caption{Structural search-space accounting. The reduction is not an end-to-end speedup.}
\label{tab:pool_cost_bill}
\scriptsize
\setlength{\tabcolsep}{3.2pt}
\begin{tabular}{lrrr}
\toprule
Quantity & Asset-specific & Market-shared & Reduction \\
\midrule
Shared strategy pools & 744 & 60 & 91.9\% \\
Active members & 3,720 & 300 & 91.9\% \\
Candidate budget & $744GN_{\rm trial}$ & $60GN_{\rm trial}$ & 91.9\% \\
\bottomrule
\end{tabular}
\end{table}

Here $G$ is the number of valid schemas considered per pool and $N_{\rm trial}=64$ is the BO trial budget. The reduction in shared strategy pools and active members is exact. End-to-end measurements in Table~\ref{tab:resource_results} additionally include schema validation, data loading, orchestration, residual fitting, and evaluation overhead.

Candidate influence is checked through
\begin{equation}
\Delta_{u,b}(x)=J_{\rm pool}(x;\mathcal{B}_b,m,k,r)-J_{\rm pool}^{(-u)}(x;\mathcal{B}_b,m,k,r),
\end{equation}
where the second term removes asset $u$ from the same block and renormalizes the remaining assets in that market portfolio. We record the complete influence distribution and the largest absolute asset share before admitting a market-shared candidate.

To compare residual magnitudes across heterogeneous parameter scales, let $\mathcal{A}_{s_j}$ be the active numeric parameters of schema $s_j$ and define
\begin{equation}
\rho_{u,j}=
\left[
\frac{1}{|\mathcal{A}_{s_j}|}
\sum_{d\in\mathcal{A}_{s_j}}
\left(
\frac{(D_{s_j}\delta_{u,j})_d}
     {\theta^{\max}_{s_j,d}-\theta^{\min}_{s_j,d}}
\right)^2
\right]^{1/2}.
\label{eq:residual_magnitude}
\end{equation}
A component is counted as boundary-saturated when either the residual bound or the projected schema bound is active within the audit tolerance. Low, middle, and high residual-magnitude groups are validation-defined tertiles of $\rho_{u,j}$ and are frozen before holdout scoring.

\begin{table}[h]
\centering
\caption{Prespecified residual and influence diagnostics. Magnitude groups are validation-defined residual-norm tertiles frozen before holdout scoring.}
\label{tab:residual_diagnostics}
\scriptsize
\setlength{\tabcolsep}{3.0pt}
\begin{tabularx}{\columnwidth}{@{}l X l@{}}
\toprule
Item & Reported statistic & Value \\
\midrule
\multicolumn{3}{@{}l}{\textit{A. Residual and influence distribution}} \\
Residual norm $\rho$ & median / P90 / maximum & 0.05 / 0.14 / 0.21 \\
Boundary saturation & fraction of active components clipped & $10\%$ \\
Boundary saturation breakdown & clipped fraction by schema and market & \shortstack{schema 6--14\%;\\market 11/9/8/10/12\%} \\
Leave-one-asset influence & P95 / maximum $|\Delta_{u,b}|$ & 0.08 / 0.21 \\
\midrule
\multicolumn{3}{@{}l}{\textit{B. Holdout performance by residual magnitude}} \\
Low tertile & assets / TR / SR / MDD & $20$ / $11\%$ / $1.2$ / $-14\%$ \\
Middle tertile & assets / TR / SR / MDD & $21$ / $14\%$ / $0.9$ / $-11\%$ \\
High tertile & assets / TR / SR / MDD & $21$ / $13\%$ / $1.1$ / $-13\%$ \\
High-tertile concentration & top-asset share / volatility-adjusted check & \shortstack{$14\%$ / SR 1.03;\\MDD $-13.2\%$} \\
\bottomrule
\end{tabularx}
\end{table}

Let $D_{m,k,r}(x)$ denote the regime-specific candidate distribution and $D_{m,k}(x)$ its market--expert parent. With $B_{\min}=20$ effective blocks, sparse-cell shrinkage is
\begin{equation}
\begin{aligned}
\widetilde D_{m,k,r}(x)
  &=\omega_{m,r}D_{m,k,r}(x)+(1-\omega_{m,r})D_{m,k}(x),\\
\omega_{m,r}
  &=\min\!\left(1,\frac{B_{m,r}}{B_{\min}}\right).
\end{aligned}
\label{eq:regime_shrinkage}
\end{equation}
The effective-block count, shrinkage weight, and pre/post-shrinkage quantiles are retained for every update. This avoids ranking a sparse regime cell by an unstable empirical 5\% quantile.

\subsection{Schema Fields}

\begin{table*}[h]
\centering
\caption{Machine-readable strategy schema and validation constraints.}
\label{tab:schema_fields_appendix}
\small
\begin{tabularx}{\textwidth}{lXX}
\toprule
Field & Constraint & Example role \\
\midrule
\texttt{expert} & trend/reversal/breakout/exposure & fixes the admissible grammar \\
\texttt{indicators} & declared list with bounded causal windows & MA, RSI, ATR, volume \\
\texttt{entry}/\texttt{exit} & typed Boolean expression tree & crossover or threshold rule \\
\texttt{position} & bounded target-exposure rule & long/flat/short or target volatility \\
\texttt{risk} & stop, leverage, cash, and turnover bounds & execution-safe constraints \\
\texttt{parameters} & typed range, integer flag, and default & search domain $\Theta_s$ \\
\texttt{provenance} & adapter, prompt, seed, source cutoff & reproducibility and leakage audit \\
\bottomrule
\end{tabularx}
\end{table*}

\section{RABO and DF-RABO Details}
\label{app:rabo_details}

\subsection{Robustness Score and Controls}

For candidate set $\Omega_{m,k,r,\tau}$, every component in Eq.~\eqref{eq:rabo} is ranked within the same pool update. $R_P$ ranks $P(J_{\rm pool}>J_{\rm bench})$; $R_{Q^{\rm cand,val}_5}$ ranks the 5\% quantile of the shrunk validation distribution; $R_{\rm med}$ ranks its median; $R_{\rm stab}$ measures performance retention under local parameter perturbations; and $R_{\rm turn}$ ranks realized turnover. Rank normalization removes incompatible scales without converting all negative lower-tail outcomes to the same clipped value.

Let $x^{\rm seed}_{m,k,r}$ denote the predeclared fixed-seed strategy-pool reference for market $m$, expert $k$, and regime $r$. Its matched-block benchmark is
\begin{equation}
J_{{\rm bench},b}^{(m,k,r)}
  =J_{\rm pool}\!\left(x^{\rm seed}_{m,k,r};\mathcal{B}_b,m,k,r\right).
\label{eq:jbench}
\end{equation}
For candidate $x$, benchmark exceedance is the fraction of the same validation blocks on which its after-cost objective is larger:
\begin{equation}
P(J_{\rm pool}>J_{\rm bench})
  =\frac{1}{B_{m,r}}\sum_{b=1}^{B_{m,r}}
  \mathbb{I}\!\left[
  J_{\rm pool}(x;\mathcal{B}_b,m,k,r)>J_{{\rm bench},b}^{(m,k,r)}
  \right].
\label{eq:benchmark_exceedance}
\end{equation}
The reference is fixed before BO and RABO, uses the same synchronized portfolio construction and cost ledger as the candidate, and is never reselected candidate by candidate.

For candidate $x=(s,\theta)$, define its scalar local-validation objective as the median after-cost outcome over the matched blocks:
\begin{equation}
J_{\rm val}(s,\theta)
  =\operatorname*{median}_{1\le b\le B_{m,r}}
  J_{\rm pool}\!\left((s,\theta);\mathcal{B}_b,m,k,r\right).
\label{eq:local_validation_objective}
\end{equation}
Let $\mathcal{A}_s$ index the active numeric parameters of schema $s$, $e_d$ be the unit vector for dimension $d$, and $[\theta_d^{\min},\theta_d^{\max}]$ its declared search range. The one-at-a-time 5\% neighborhood is
\begin{equation}
\mathcal{N}(\theta)=
\left(
\bigcup_{d\in\mathcal{A}_s}
\left\{
\Pi_{\Theta_s}\!\left(\theta\pm.05(\theta_d^{\max}-\theta_d^{\min})e_d\right)
\right\}
\right)\setminus\{\theta\}.
\label{eq:stability_neighborhood}
\end{equation}
Projection enforces the declared bounds; integer parameters are rounded to the nearest legal integer, projected duplicates are replaced by the nearest distinct legal neighbor when one exists, and categorical operators remain fixed. After duplicate neighbors are removed, local stability is
\begin{equation}
\operatorname{Stability}(\theta)=
\operatorname{clip}_{[0,1]}\!\left(
1-\operatorname*{median}_{\theta'\in\mathcal{N}(\theta)}
\frac{\left|J_{\rm val}(s,\theta')-J_{\rm val}(s,\theta)\right|}
{\left|J_{\rm val}(s,\theta)\right|+\varepsilon_J}
\right).
\label{eq:local_stability}
\end{equation}

\begin{algorithm}[h]
\caption{Market-shared RABO pool update}
\label{alg:rabo_appendix}
\begin{algorithmic}[1]
\REQUIRE market $m$, expert $k$, regime $r$, causal cutoff $\tau$
\STATE Form grammar-valid schemas using records visible by $\tau$
\STATE Run BO over each declared parameter space
\FOR{candidate $x_i$}
  \STATE Evaluate synchronized $J_{\rm pool}(x_i)$ on purged validation blocks
  \STATE Apply Eq.~\eqref{eq:regime_shrinkage} when the regime cell is sparse
  \STATE Compute exceedance, $Q^{\rm cand,val}_5$, median, stability, and turnover ranks
  \STATE Compute $S_{\rm RABO}(x_i)$ using Eq.~\eqref{eq:rabo}
\ENDFOR
\STATE Fully evaluate every candidate eligible for admission
\STATE Admit the five highest audited candidates and fit bounded asset-specific parameter adjustments
\RETURN $\mathcal{P}_{m,k,r}$ and pool-level robustness evidence $c^{\rm rob}_{m,r,k}$
\end{algorithmic}
\end{algorithm}

\begin{table*}[h]
\centering
\caption{Five-seed mean RABO-control results on the frozen holdout under the matched execution protocol.}
\label{tab:rabo_controls_appendix}
\scriptsize
\setlength{\tabcolsep}{3.2pt}
\begin{tabularx}{\textwidth}{lXrrrrr}
\toprule
Control & Admission rule & TR\% & SR & MDD\% & $Q^{\rm port}_{5,20d}$\% & Turn. \\
\midrule
BO-best & highest validation peak only & 15.0 & 1.02 & -17.3 & -11.85 & 9.8 \\
Median-only & median of $\widetilde D_{m,k,r}(x)$ & 13.5 & 1.18 & -14.8 & -9.10 & 8.0 \\
Lower-tail-only & $Q^{\rm cand,val}_5$; CVaR sensitivity retained & 13.2 & 1.14 & -14.2 & -8.20 & 7.5 \\
Stability-only & local parameter-neighborhood stability & 12.9 & 1.10 & -14.6 & -9.35 & 7.0 \\
Turnover-matched BO & BO-best after matching realized turnover to RABO & 14.3 & .99 & -16.2 & -9.87 & 7.3 \\
Equal-weight audit & equal signed weights for the five RABO components & 13.6 & 1.20 & -14.5 & -9.59 & 7.6 \\
Full RABO & Eq.~\eqref{eq:rabo} with validation-frozen weights & 14.1 & 1.33 & -13.7 & -8.22 & 7.2 \\
Weight sensitivity A & $(.35,.25,.15,.15,-.10)$ & 14.0 & 1.31 & -14.0 & -8.35 & 7.3 \\
Weight sensitivity B & $(.25,.35,.15,.15,-.10)$ & 13.9 & 1.32 & -13.9 & -8.15 & 7.1 \\
Weight sensitivity C & $(.30,.30,.20,.10,-.10)$ & 14.2 & 1.29 & -14.1 & -8.44 & 7.4 \\
\bottomrule
\end{tabularx}
\end{table*}

\subsection{Mixed Representation and Fidelity Gate}

The evaluator encoder produces a frozen schema embedding $e_s\in\R^{d_s}$. For the union of expert-specific parameter domains, $\tilde\theta\in[0,1]^{d_\theta}$ stores normalized values and $a_s\in\{0,1\}^{d_\theta}$ marks active dimensions. Inactive values are ignored by the numerical distance, integer dimensions are rounded only before execution, and categorical choices use separate embeddings.

The semantic and masked numerical kernels are combined with a paired Kennedy--O'Hagan fidelity model \cite{kennedy2000predicting}:
\begin{equation}
\begin{aligned}
k(x,x')&=k_{\rm sem}(e_s,e_{s'})\,k_{\rm num}([\tilde\theta;a_s],[\tilde\theta';a_{s'}]),\\
J_R(x)&=\rho J_L(x)+\delta(x),\qquad
p_\delta(x)=\frac{\sigma_\delta^2(x)}{\rho^2\sigma_L^2(x)+\sigma_\delta^2(x)}.
\end{aligned}
\end{equation}
Candidates with high discrepancy dominance receive priority for full evaluation; every admission candidate still receives the same complete after-cost walk-forward audit. Table~\ref{tab:dfrabo} reports a separate budget-matched six-method search diagnostic comparing vanilla BO, full-fidelity RABO search, LLM-initialized BO, evaluator-score-only search, random-fidelity RABO, and DF-RABO. Its metrics belong to that diagnostic benchmark and are not directly comparable to the end-to-end equal-market holdout ablation in Table~\ref{tab:combined_ablation}.

\begin{table*}[t]
\centering
\caption{Independent dual-fidelity search diagnostic under matched full-evaluation budgets. Its metrics are not directly comparable to the end-to-end equal-market holdout ablation in Table~\ref{tab:combined_ablation}. Elapsed time is measured in the same environment and normalized to vanilla BO.}
\label{tab:dfrabo}
\scriptsize
\setlength{\tabcolsep}{4.2pt}
\begin{tabular}{lrrrrrr}
\toprule
Method & SR & MDD\% & $Q^{\rm port}_{5,20d}$\% & Full backtests & Evaluator calls & Relative time \\
\midrule
Vanilla BO & .69 & -17.2 & -4.31 & 64 & 0 & 1.00$\times$ \\
Full-fidelity RABO search & .81 & -13.8 & -2.74 & 64 & 0 & 1.07$\times$ \\
LLM-initialized BO & .74 & -16.0 & -3.65 & 64 & 20 & 1.05$\times$ \\
Evaluator-score-only & .40 & -21.5 & -5.92 & 0 & 200 & .18$\times$ \\
Random-fidelity RABO & .57 & -17.0 & -4.10 & 28 & 200 & .55$\times$ \\
\textbf{DF-RABO} & \textbf{.82} & \textbf{-13.6} & \textbf{-2.69} & \textbf{28} & \textbf{200} & \textbf{.56$\times$} \\
\bottomrule
\end{tabular}
\end{table*}

\section{Training and Executed-Credit Accounting}
\label{app:training}

The first training stage minimizes regime cross-entropy on training records. The second stage freezes the backbone and optimizes the router, expert heads, and execution adapters with clipped actor--critic updates. The reward includes after-cost return and penalties for drawdown, turnover, constraint projection, and abrupt router changes. The validation objective and all penalty weights are fixed before the frozen trading holdout.

For synchronized raw expert vector $v_{t+1\mid t,m,k}$, define smoothed sign-specific shares for each asset component $u$:
\begin{equation}
a^{\pm}_{t+1\mid t,u,k}=
\frac{[\pm v_{t+1\mid t,m,k,u}]_+ + \epsilon/K}
{\sum_j[\pm v_{t+1\mid t,m,j,u}]_+ + \epsilon}.
\end{equation}
The executed allocation is
\begin{equation}
\tilde h_{t+1\mid t,m,k,u}=
\begin{cases}
a^+_{t+1\mid t,u,k}h^*_{t+1\mid t,m,u}, & h^*_{t+1\mid t,m,u}\ge0,\\
a^-_{t+1\mid t,u,k}h^*_{t+1\mid t,m,u}, & h^*_{t+1\mid t,m,u}<0.
\end{cases}
\end{equation}
Because $\sum_k a^{\pm}_{t+1\mid t,u,k}=1$, expert allocations reconcile exactly with the executed portfolio. Cost is allocated by each expert's post-projection turnover, not its unexecuted raw target. The leave-one-expert-out diagnostic independently recomputes projection and settlement without expert $k$ and compares the counterfactual change with the allocated contribution.

\section{Execution Rules and Additional Evaluation}
\label{app:execution}

\begin{table*}[h]
\centering
\caption{Market-specific feasibility, cost components, and operational handling.}
\label{tab:execution_rules_appendix}
\footnotesize
\renewcommand{\arraystretch}{0.92}
\begin{tabularx}{\textwidth}{lXXXX}
\toprule
Market & Position feasibility & Rebalance/calendar & Applicable cost components & Additional handling \\
\midrule
A-shares & long-only, asset/gross caps & exchange days, T+1 & commission, sell stamp duty, slippage & adjusted prices and suspensions \\
U.S. equities & bounded long/short, margin & exchange calendar & commission, spread/impact proxy & corporate actions and borrow limits \\
ETFs & venue-dependent long/short & venue calendar & commission and slippage & distributions and underlying class \\
Futures & contract leverage and margin & lagged dominant contract; next-open roll & fees, slippage, roll cost & multipliers and roll ledger \\
Cryptocurrency & bounded leverage & continuous 24/7 & fees, slippage, funding & exchange source and outage checks \\
\bottomrule
\end{tabularx}
\end{table*}

\subsection{Projection and Priority Clipping}

Equation~\eqref{eq:projection} is a convex projection with an $\ell_1$ turnover term; an exact epigraph-form QP solution is retained as an audit reference. The implemented fast path applies the same predeclared priority order to every compatible method. Each stage can only reduce the requested exposure, and the final ledger rejects any holding that fails the complete feasible-set check.

\begin{algorithm}[h]
\caption{Market-level priority clipping}
\label{alg:priority_clipping}
\begin{algorithmic}[1]
\REQUIRE raw target $h^{\rm raw}_{t+1\mid t,m}$, prior holding $h^*_{t,m}$, rules $\mathcal{C}_m$
\STATE Remove unavailable orders and enforce long/short eligibility
\STATE Clip each asset to its position bound
\STATE Enforce gross/net exposure and leverage caps
\STATE Enforce the required cash buffer
\STATE Clip $q=h-h^*_{t,m}$ to the turnover limit
\STATE Recheck $\mathcal{C}_m$; reduce exposure if needed
\RETURN feasible $h^*_{t+1\mid t,m}$ and order $q_{t+1,m}$
\end{algorithmic}
\end{algorithm}

At date $t$, the model uses records available through the market close and submits $q_{t+1,m}=h^*_{t+1\mid t,m}-h^*_{t,m}$ at the next tradable open. Suspended or unavailable assets retain their prior feasible holding until an order can be executed. The primary normalized ledger applies Equation~\eqref{eq:normalized_cost} as one all-in charge. Its component fields allocate that total for attribution only. In the market-specific evaluation, commission, slippage, stamp duty, roll cost, and funding rules replace---and never supplement---the normalized charge; A-share stamp duty is nonzero only for executed sells.

\subsection{Metric Definitions and Equal-Market Aggregation}

For one fixed training seed, let $r^{\rm net}_{t,m}$ be the daily after-cost return of market portfolio $m$ over its $T_m$ native sessions; the seed index is suppressed. The daily risk-free rate is fixed at zero. Define $A_m=252$ for exchange-traded business-day markets and $A_m=365$ for cryptocurrency. With
\begin{align}
\bar r^{\rm net}_m
  &=\frac{1}{T_m}\sum_{t=1}^{T_m}r^{\rm net}_{t,m}, \notag\\
(s^{\rm net}_m)^2
  &=\frac{1}{T_m-1}\sum_{t=1}^{T_m}
    \left(r^{\rm net}_{t,m}-\bar r^{\rm net}_m\right)^2, \notag\\
\operatorname{SR}_m
  &=\sqrt{A_m}\,\frac{\bar r^{\rm net}_m}{s^{\rm net}_m},
\label{eq:sharpe_metric}
\end{align}
the reported Sharpe ratio is therefore computed from the market portfolio, not by averaging asset-level Sharpe ratios. Let $V_{0,m}=1$ and $V_{t,m}=\prod_{\tau=1}^{t}(1+r^{\rm net}_{\tau,m})$. Maximum drawdown is
\begin{equation}
\operatorname{MDD}_m
  =\min_{1\le t\le T_m}
   \left(\frac{V_{t,m}}{\max_{0\le s\le t}V_{s,m}}-1\right)\le 0,
\label{eq:mdd_metric}
\end{equation}
and is reported as a negative percentage. Annualized one-way turnover uses the same executed orders as the cost ledger:
\begin{equation}
\operatorname{Turn}_m
  =\frac{A_m}{T_m}\sum_{t=1}^{T_m}\lVert q_{t,m}\rVert_1.
\label{eq:turnover_metric}
\end{equation}
Finally, after computing any reported metric $g_m$ on each market's native calendar, the equal-market summary is
\begin{equation}
g^{\rm eq}=\frac{1}{|\mathcal{M}|}\sum_{m\in\mathcal{M}}g_m
          =\frac{1}{5}\sum_{m\in\mathcal{M}}g_m.
\label{eq:equal_market_metric}
\end{equation}
This is a reporting operator over five separately settled market portfolios, not a globally synchronized fund.

\begin{figure}[h]
\centering
\includegraphics[width=0.92\linewidth]{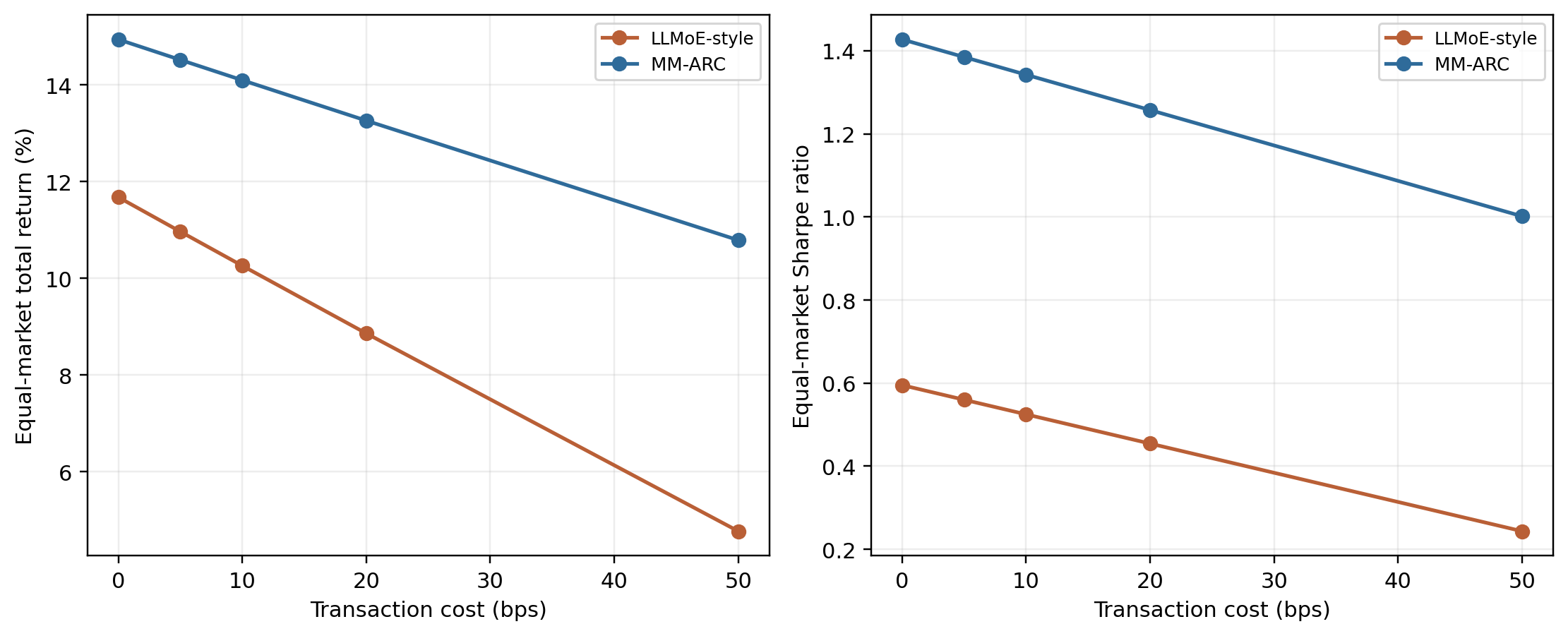}
\Description{After-cost portfolio performance is plotted over the predeclared transaction-cost grid.}
\caption{Transaction-cost sensitivity over 0, 5, 10, 20, and 50 bps. Predictions, gross returns, and orders remain fixed; only the charged friction changes.}
\label{fig:cost_sensitivity_appendix}
\end{figure}

\begin{table}[t]
\centering
\caption{Market-specific friction sensitivity. This setting replaces the normalized charge; predictions, gross returns, and orders are fixed. Rates are predeclared execution assumptions, and entries are five-seed means.}
\label{tab:market_specific_costs}
\small
\setlength{\tabcolsep}{3.5pt}
\begin{tabular}{lrrrrrrr}
\toprule
Setting & Cost bps & \multicolumn{3}{c}{MM-ARC} & \multicolumn{3}{c}{LLMoE-style} \\
 & & TR\% & SR & MDD\% & TR\% & SR & MDD\% \\
\midrule
A-shares & 18 & 15.0 & 1.40 & -12.6 & 12.2 & 0.64 & -16.7 \\
U.S. Equities & 8 & 24.7 & 1.55 & -16.6 & 19.8 & 0.68 & -22.2 \\
ETFs & 6 & 15.1 & 1.50 & -9.8 & 10.6 & 0.58 & -15.2 \\
Futures & 5 & 9.6 & 1.37 & -9.6 & 6.4 & 0.49 & -13.4 \\
Crypto & 25 & 5.0 & 0.74 & -20.3 & 0.4 & 0.10 & -22.7 \\
\midrule
Equal-market & mixed & 13.9 & 1.31 & -13.8 & 9.9 & 0.50 & -18.0 \\
\bottomrule
\end{tabular}
\end{table}

\section{Complete Numeric Diagnostics}
\label{app:numeric_diagnostics}

Table~\ref{tab:per_seed_results} reports all five matched equal-market runs used for the main mean and SD. Learned methods use training seeds, stochastic technical search uses TPE/BO seeds, and Buy-and-Hold is invariant. Table~\ref{tab:seed_stability_statistics} retains auxiliary seed dispersion, marginal intervals, turnover, and Holm tests separately from Table~\ref{tab:statistical_summary}. Table~\ref{tab:block_length_sensitivity} reports the 10/20/40-session path-level sensitivity analysis, and Table~\ref{tab:per_market_audit} exposes the per-market results. Tables~\ref{tab:regime_coverage} and \ref{tab:numeric_integrity} report unresampled regime support and record-level checks.

\begin{table*}[t]
\centering
\caption{Equal-market results for five matched runs under the normalized 10-bps (0.10\%) one-way all-in cost. Each cell is TR\% / SR / MDD\%. Seeds index training runs for learned methods and TPE/BO runs for stochastic technical search; B\&H is identical because neither changes its observed market path.}
\label{tab:per_seed_results}
\scriptsize
\setlength{\tabcolsep}{4.0pt}
\begin{tabular}{lccccc}
\toprule
Method & Seed 42 & Seed 43 & Seed 44 & Seed 45 & Seed 46 \\
\midrule
B\&H & 14.0/0.58/-24.7 & 14.0/0.58/-24.7 & 14.0/0.58/-24.7 & 14.0/0.58/-24.7 & 14.0/0.58/-24.7 \\
Stoch. technical search & 7.6/0.49/-20.2 & 9.9/0.44/-19.8 & 8.9/0.41/-20.3 & 9.5/0.47/-19.9 & 8.8/0.44/-21.2 \\
PPO & 7.9/0.39/-18.6 & 9.2/0.54/-18.2 & 7.8/0.39/-17.9 & 8.7/0.47/-18.3 & 8.2/0.44/-18.3 \\
Time-VLM-style & 8.6/0.45/-18.5 & 10.8/0.59/-18.4 & 9.2/0.50/-19.3 & 10.2/0.49/-17.8 & 9.7/0.44/-18.7 \\
FinAgent (adapted) & 7.5/0.41/-20.7 & 9.3/0.49/-20.4 & 8.4/0.42/-20.8 & 9.5/0.46/-20.2 & 8.8/0.43/-20.8 \\
LLMoE-style & 9.2/0.46/-18.3 & 11.4/0.59/-17.8 & 9.8/0.50/-18.9 & 10.7/0.55/-18.2 & 10.8/0.56/-18.2 \\
\textbf{MM-ARC} & \textbf{13.0/1.28/-13.9} & \textbf{15.1/1.33/-13.8} & \textbf{13.4/1.28/-13.9} & \textbf{14.8/1.36/-13.7} & \textbf{14.2/1.41/-13.3} \\
\bottomrule
\end{tabular}
\end{table*}

\begin{table*}[t]
\centering
\caption{Optimization-stability and marginal method summaries for the frozen trading holdout. Mean and Seed SD summarize five independently trained equal-market seed aggregates. Marginal time-block CIs describe each method separately and are not substitutes for the paired difference CI in Table~\ref{tab:statistical_summary}. Holm $p$ values are auxiliary optimization-stability evidence from two-sided tests pairing the five seed aggregates against MM-ARC.}
\label{tab:seed_stability_statistics}
\scriptsize
\setlength{\tabcolsep}{4.2pt}
\renewcommand{\arraystretch}{1.08}
\begin{tabular}{lrrrrrrr}
\toprule
Method & Mean SR & \shortstack{SR\\Seed SD} & \shortstack{Marginal block\\95\% CI (SR)} & \shortstack{Mean\\MDD\%} & \shortstack{MDD\\Seed SD} & \shortstack{One-way turnover\\($\times$/yr)} & Holm $p$ \\
\midrule
PPO & 0.45 & 0.06 & [-0.03, 1.31] & -18.3 & 0.3 & 11.1 & $<.001$ \\
FinAgent (adapted) & 0.44 & 0.03 & [-0.01, 0.94] & -20.6 & 0.3 & 17.5 & $<.001$ \\
Time-VLM-style & 0.49 & 0.06 & [0.02, 1.17] & -18.5 & 0.5 & 13.6 & $<.001$ \\
LLMoE-style & 0.53 & 0.05 & [0.08, 1.19] & -18.3 & 0.4 & 12.5 & $<.001$ \\
\textbf{MM-ARC} & \textbf{1.33} & \textbf{0.06} & \textbf{[0.87, 2.69]} & \textbf{-13.7} & \textbf{0.2} & \textbf{7.2} & --- \\
\bottomrule
\end{tabular}
\end{table*}

\begin{table*}[t]
\centering
\caption{Block-length sensitivity for paired path-level inference. Each metric cell reports the point difference followed by its paired 95\% CI from the 1,200-draw market-stratified circular-block bootstrap.}
\label{tab:block_length_sensitivity}
\scriptsize
\setlength{\tabcolsep}{4.0pt}
\renewcommand{\arraystretch}{1.08}
\begin{tabular}{llccc}
\toprule
Comparison & Block
& \shortstack{$\Delta$SR\\{[}95\% CI{]}}
& \shortstack{$\Delta$MDD (pp)\\{[}95\% CI{]}}
& \shortstack{$\Delta Q^{\rm port}_{5,20d}$ (pp)\\{[}95\% CI{]}} \\
\midrule
\multirow{3}{*}{MM-ARC $-$ LLMoE-style}
 & 10 & \shortstack{$+0.80$\\$[0.26,1.46]$} & \shortstack{$+4.6$\\$[2.03,6.94]$} & \shortstack{$+2.15$\\$[.78,3.56]$} \\
 & 20 & \shortstack{$+0.80$\\$[.13,1.52]$} & \shortstack{$+4.6$\\$[1.88,7.15]$} & \shortstack{$+2.15$\\$[.47,3.85]$} \\
 & 40 & \shortstack{$+0.80$\\$[.14,1.59]$} & \shortstack{$+4.6$\\$[1.24,7.79]$} & \shortstack{$+2.15$\\$[.33,4.07]$} \\
\midrule
\multirow{3}{*}{MM-ARC $-$ market-specific static}
 & 10 & \shortstack{$+0.12$\\$[.06,.21]$} & \shortstack{$+0.8$\\$[.05,1.32]$} & \shortstack{$+0.63$\\$[.15,1.14]$} \\
 & 20 & \shortstack{$+0.12$\\$[.02,.26]$} & \shortstack{$+0.8$\\$[.11,1.42]$} & \shortstack{$+0.63$\\$[.03,1.33]$} \\
 & 40 & \shortstack{$+0.12$\\$[.07,.33]$} & \shortstack{$+0.8$\\$[.15,1.52]$} & \shortstack{$+0.63$\\$[.04,1.44]$} \\
\midrule
\multirow{3}{*}{Full RABO $-$ BO-best}
 & 10 & \shortstack{$+0.31$\\$[.13,.55]$} & \shortstack{$+3.6$\\$[1.36,5.68]$} & \shortstack{$+3.63$\\$[1.64,5.65]$} \\
 & 20 & \shortstack{$+0.31$\\$[.03,.62]$} & \shortstack{$+3.6$\\$[1.16,5.88]$} & \shortstack{$+3.63$\\$[1.21,6.07]$} \\
 & 40 & \shortstack{$+0.31$\\$[.06,.73]$} & \shortstack{$+3.6$\\$[.62,6.37]$} & \shortstack{$+3.63$\\$[.84,6.53]$} \\
\bottomrule
\end{tabular}
\end{table*}

\begin{table*}[t]
\centering
\caption{Per-market five-seed mean results under the normalized 10-bps protocol. Each cell is TR\% / SR / MDD\%, computed from the retained method--seed--market daily-return records. Market cells are rounded separately; equal-market summaries are computed from the unrounded seed--market records and can therefore differ by one displayed rounding unit from an average of the displayed cells.}
\label{tab:per_market_audit}
\scriptsize
\setlength{\tabcolsep}{4.0pt}
\begin{tabular}{lccccc}
\toprule
Method & A-shares & U.S. equities & ETFs & Futures & Crypto \\
\midrule
Buy-and-Hold
& $19.8/.72/-22.0$
& $32.3/.92/-29.1$
& $16.1/.75/-16.7$
& $4.6/.35/-18.2$
& $-2.9/.18/-37.6$ \\
LLMoE-style
& $13.3/.69/-16.9$
& $19.7/.68/-22.5$
& $10.4/.56/-15.6$
& $5.7/.44/-13.7$
& $2.9/.28/-22.8$ \\
Market-specific learned static
& $15.2/1.32/-13.3$
& $23.5/1.42/-17.4$
& $14.2/1.30/-10.8$
& $8.8/1.12/-10.7$
& $7.3/.89/-20.3$ \\
\textbf{MM-ARC}
& $15.8/1.46/-12.5$
& \textbf{24.2/1.55/-16.5}
& $14.9/1.45/-9.8$
& $9.1/1.30/-9.6$
& $6.5$/\textbf{.94/-20.6} \\
\bottomrule
\end{tabular}
\end{table*}

\begin{table}[t]
\centering
\caption{Predicted router-regime coverage. Entries are average observed sessions per seed (share of market timestamps).}
\label{tab:regime_coverage}
\scriptsize
\setlength{\tabcolsep}{3.2pt}
\begin{tabular}{lrrr}
\toprule
Setting & Uptrend & Downtrend & Sideways \\
\midrule
A-shares & 157.8 (65.2\%) & 38.2 (15.8\%) & 46.0 (19.0\%) \\
U.S. Equities & 151.8 (60.5\%) & 45.4 (18.1\%) & 53.8 (21.4\%) \\
ETFs & 106.4 (42.4\%) & 68.6 (27.3\%) & 76.0 (30.3\%) \\
Futures & 118.0 (47.0\%) & 45.4 (18.1\%) & 87.6 (34.9\%) \\
Crypto & 133.0 (36.4\%) & 29.2 (8.0\%) & 202.8 (55.6\%) \\
\bottomrule
\end{tabular}
\end{table}

\begin{table}[t]
\centering
\caption{Numerical integrity checks on daily records from the frozen trading holdout.}
\label{tab:numeric_integrity}
\small
\resizebox{\columnwidth}{!}{%
\begin{tabular}{lr}
\toprule
Identity & Maximum absolute error \\
\midrule
Gross return $-$ costs $-$ net return & $<7\times10^{-18}$ \\
Sum of component costs $-$ total cost & $<7\times10^{-18}$ \\
Previous holding $+$ order $-$ executed holding & $<7\times10^{-18}$ \\
Instrument aggregation $-$ market portfolio & $<6\times10^{-17}$ \\
\bottomrule
\end{tabular}%
}
\end{table}

\FloatBarrier

\section{Result Provenance and Consistency}
\label{app:result_provenance}

The anonymous repository contains an executable pipeline and related experiment results for \method{}. It covers data processing, training, strategy search, evaluation, and report generation. Full \method{} values reconcile across the main results, ablations, RABO comparison, per-seed report, and the normalized all-in 10-bps cost point whenever their aggregation levels match.

Metrics are computed on each seed--market portfolio before equal-market averaging. No asset-average Sharpe ratio is substituted for portfolio Sharpe. Turnover is the absolute change in executed holdings, and costs use the same order ledger that reconciles consecutive holdings. Candidate-level $Q^{\rm cand,val}_5$ remains confined to past validation blocks, whereas the reported $Q^{\rm port}_{5,20d}$ is computed from rolling 20-session after-cost returns in the corresponding evaluation record.

The release scope is the executable pipeline and its accompanying experiment results. It does not claim to redistribute the complete upstream market-data corpus or every intermediate record produced during the research process. The favorable paired results and family-level rejections are both reported in this paper.

\end{document}